%% file: nn_decoder01.tex
\documentclass[a4paper,twocolumn,superscriptaddress,11pt,accepted=2018-05-04]{quantumarticle}
\pdfoutput=1
\usepackage[utf8]{inputenc}
\usepackage[english]{babel}
\usepackage{amsmath}
\usepackage{amsfonts}
\usepackage{amssymb}
\usepackage{graphicx}
\usepackage{xcolor-solarized}
\usepackage{tikz,pgfplots}
\usetikzlibrary{matrix,chains,positioning,angles,quotes,calc,arrows,3d,decorations.pathreplacing}
\usepackage{amsthm}
\usepackage{color}
\usepackage[subpreambles=true]{standalone}
\usepackage{subcaption}
\usepackage[numbers,sort&compress]{natbib}
\usepackage{textcomp}

\newcommand\blfootnote[1]{%
	\begingroup
	\renewcommand\thefootnote{}\footnote{#1}%
	\addtocounter{footnote}{-1}%
	\endgroup
}

\newtheorem{observation}{Observation}
\newcommand{\plog}{\bar{P}}
\newcommand{\real}{\mathbb{R}}

\newcommand{\aref}[1]{\hyperref[#1]{Appendix~\ref{#1}}}

\title{Scalable Neural Network Decoders\\ for Higher Dimensional Quantum Codes}
\author{N.~P.~Breuckmann}
\affiliation{Institute for Quantum Information, RWTH Aachen University, Germany}

\author{X.~Ni}
\affiliation{Institute for Quantum Information, RWTH Aachen University, Germany}
\affiliation{Max Planck Institute of Quantum Optics, Germany}

\date{\today}

\begin{document}
\maketitle
\blfootnote{The authors contribute equally to this work.}

\input{abstract}

\section{Introduction}
\label{sec:Intro}
\input{introduction}


\section{Previous Work}
\label{sec:previous}
\input{previous_work}

\section{The 4D Toric Code}
\label{sec:4DTC}
\input{4dtc}

\section{Basics of Machine Learning}
\label{sec:ML}
\input{machine_learning}

\section{Machine Learning for Decoding Quantum Codes}
\label{sec:ML_for_decoding}
\input{machine_learning_for_decoding}

\section{Numerical Analysis of the Neural Network Decoder}
\label{sec:Num}
\input{numerical_analysis}

\section{Discussion}
\input{discussion}


\bibliographystyle{plainnat}
\bibliography{bibliography} 

\appendix

\section{Training of Neural networks}
\input{training_neural_network}

\section{Description of the parallel line decoder}
\input{parallel_line_decoder}

\section{Brief summary of the AlphaGo architecture}
\input{alphago}

\section{A toy example of error correction with 1-nearest neighbor algorithm}
\input{lookup_table}

\section{Analysis of the Convolutional Neural Network}
\input{analysis_cnn}

\end{document}

%% file: abstract.tex
\begin{abstract}
Machine learning has the potential to become an important tool in quantum error correction as
it allows the decoder to adapt to the error distribution of a quantum chip.
An additional motivation for using neural networks is the fact that they can be evaluated by dedicated hardware which is very fast and consumes little power.
Machine learning has been previously applied to decode the surface code.
However, these approaches are not scalable as the training has to be redone for every system size which becomes increasingly difficult.
In this work the existence of local decoders for higher dimensional codes leads us to use a low-depth convolutional neural network to locally assign a likelihood of error on each qubit.
For noiseless syndrome measurements, numerical simulations show that the decoder has a threshold of around $7.1\%$ when applied to the 4D toric code.
When the syndrome measurements are noisy, the decoder performs better for larger code sizes when the error probability is low.
We also give theoretical and numerical analysis to show how a convolutional neural network is different from the 1-nearest neighbor algorithm, which is a baseline machine learning method.
\end{abstract}

%% file: introduction.tex
A full-featured quantum computer will rely on some form of error correction as physical qubits are prone to the effects of environmental noise.
When using error correcting codes, decoders play a large role in the performance of the fault-tolerant protocols.
Using neural networks to decode the surface code has been suggested in earlier works \cite{torlai2016neural,varsamopoulos2017decoding,baireuther2017machine,krastanov2017deep}.
The primary motivation inspiring these works to use neural networks is their ability to adapt to the error model. 
However, an issue that has not been addressed so far is scalability: 
in previous approaches, the neural networks have to be re-trained for every system size, despite the fact that in general machine learning methods become problematic when the input space becomes too large.
Thus, it is interesting to ask whether there exists a family of quantum error correcting codes, which we can decode using neural networks in a clearly scalable way.

To provide a positive answer to this question, we introduce a decoder for the four-dimensional version of the toric code based on neural networks for which the training only has to be performed once on a small system size.
Afterwards the decoder can be scaled up to arbitrary system sizes without re-training.
The layout of the network is a \emph{convolutional neural network} which are widely used as a building block in  image recognition.
Furthermore, the neural network is constant depth with respect to the scaling of the system size.
Our approach is informed by the existence of local decoders for the 4D toric code \cite{DKLP,BDDTlocaldecoders,Breuckmann2017Homological,Pthesis}.
A drawback of those decoders is their low threshold of less than $2\%$ against independent noise (assuming perfect syndrome measurements).
The (global) minimum-weight decoder, on the other hand, has a threshold of $10.1\%$~\cite{TNthreshold4D} but it is not known to be computationally efficient.
Our decoder shown a threshold of $7.1\%$ while having a complexity close to local decoders.
This result is very close to the $7.3\%$ reported in \cite{4drenormalization} which uses a renormalization approach.
When the syndrome measurements are noisy, our decoder still performs better for larger lattice when error probability is low, see \autoref{fig:memory_time} for the plot. More numerical simulation needs to be done to determine the threshold.



This paper is structured as follows.
In \autoref{sec:previous} we discuss previous work in which machine learning was applied to decode quantum codes.
 In  \autoref{sec:4DTC} we give a review of the 4D toric code and its properties. In \autoref{sec:ML}, we provide a general introduction to machine learning. 
In \autoref{sec:ML_for_decoding} we describe the application of convolutional neural networks as a subroutine in our decoding algorithm.
In \autoref{sec:Num} we discuss the results of Monte-Carlo simulations.
In \aref{appendix:training_neural_network}, we give a short introduction to the backpropagation algorithm which we use to train the neural network.
In \aref{appendix:nearest_neighbor}, we give a toy example which shows how a baseline machine learning model can fail at a task similar to decoding when the system size becomes large. This highlights the importance of choosing a translational invariant model as we do.
In \aref{appendix:analysis_larger_network}, we observed and analyzed one property of multi-layer convolutional neural networks, and show it fits nicely to the task of assigning likelihood of qubit errors for high dimensional toric code.

%% file: previous_work.tex
It is known for a while in the classical error correction community that the errors and syndromes of a parity checking code can be put together into a probabilistic graphical model, which is a major tool for machine learning. Probabilistic graphical models include for example Bayesian networks and the Ising model (see~\cite{yedidia2003understanding} for an introduction to probabilistic graphical model and connection to error correcting codes).
In~\cite{frey1998revolution} (and references therein), it is shown that by using belief propagation, the decoder can achieve very good performance for LDPC and turbo code that are close to the Shannon limit.
We want to note that while there is no formal connection (to our knowledge) between neural networks and graphical models, a graphical model of a problem is often helpful for designing the architecture of the neural network.
For example, in~\cite{nachmani2016learning} the structure of the Tanner graph is used to construct a neural network, in order to improve belief propagation for (classical) codes that have high density checks.

In~\cite{poulin2008iterative}, the authors discussed applying belief propagation to decoding of quantum stabilizer codes. While stabilizer codes and classical parity checking codes share many similarities, the authors highlighted several issues that might cause belief propagation to work much worse for stabilizer codes (even with low density checks).
In~\cite{duclos2010fast, duclos2013fault}, the authors used belief propagation as a subroutine in the renormalization decoder for topological codes, with the purpose of synchronizing marginal probabilities across neighbouring cells.

Neural network decoder for quantum stabilizer codes has been studied in~\cite{torlai2016neural,varsamopoulos2017decoding,baireuther2017machine,krastanov2017deep} (more precisely in~\cite{torlai2016neural} hidden Boltzmann machines are used instead of neural networks).
Our work will use a similar scheme as~\cite{torlai2016neural,krastanov2017deep}, where the neural networks predict which qubits have undergone an error.
In~\cite{varsamopoulos2017decoding,baireuther2017machine}, the neural networks output 1 or 2 bits which will correct the final measurement of the logical operators.
The main difference of our work is that our machine learning decoder is naturally scalable through the use of convolutional neural networks.

%% file: 4dtc.tex
\subsection{Definition}
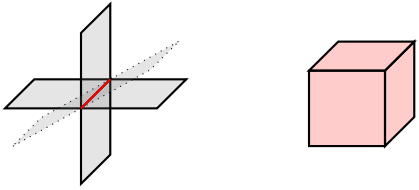
\begin{figure}
\begin{center}
\begin{tikzpicture}
\pgfmathsetmacro{\L}{1}
\draw[ thick, black,fill=gray, fill opacity=0.2] (-4,-0.5,0) -- ++(-\L,0,0) -- ++(0,0,-\L) -- ++(\L,0,0) -- cycle;
\draw[ thick, black,fill=gray, fill opacity=0.2] (-4,-0.5,0) -- ++(\L,0,0) -- ++(0,0,-\L) -- ++(-\L,0,0);
\draw[ thick, black,fill=gray, fill opacity=0.2] (-4,-0.5,0) -- ++(0,-\L,0)  -- ++(0,0,-\L) -- ++(0,\L,0) -- ++(0,0,\L);
\draw[ thick, black,fill=gray, fill opacity=0.2] (-4,-0.5,0) -- ++(0,\L,0) -- ++(0,0,-\L) -- ++(0,-\L,0);
\draw[ dotted, black,fill=gray, fill opacity=0.2] (-4,-0.5,0) -- ++(0.9\L,0.5\L,0) -- ++(0,0,-\L) -- ++(-0.9\L,-0.5\L,0) -- cycle;
\draw[ dotted, black,fill=gray, fill opacity=0.2] (-4,-0.5,0) -- ++(-0.9\L,-0.5\L,0) -- ++(0,0,-\L) -- ++(0.9\L,0.5\L,0) -- cycle;
\draw[ thick, red] (-4,-0.5,0) -- ++(0,0,-\L);
\draw[ thick, black,fill=red, fill opacity=0.2] (0,0,0) -- ++(-\L,0,0) -- ++(0,-\L,0) -- ++(\L,0,0) -- cycle;
\draw[ thick, black,fill=red, fill opacity=0.2] (0,0,0) -- ++(0,0,-\L) -- ++(0,-\L,0) -- ++(0,0,\L) -- cycle;
\draw[ thick, black,fill=red, fill opacity=0.2] (0,0,0) -- ++(-\L,0,0) -- ++(0,0,-\L) -- ++(\L,0,0) -- cycle;
\end{tikzpicture}
\end{center}
\caption{The stabilizer checks of the 4D toric code correspond to edges which act as Pauli-$X$ on all qubits incident to an edge (left) and cubes which act as Pauli-$Z$ on all qubits incident to an edge (right).}
\label{fig:4DTCstabilizer}
\end{figure}

The {\em 4D toric code} is a stabilizer quantum code defined on a four-dimensional (4D) hypercubic lattice which was first considered in \cite{DKLP}.
We will consider periodic boundary conditions so that topologically the lattice is a 4D torus.
The faces of the lattice are squares and they are identified with the qubits.
From now on we will use the words {\em face} and {\em qubit} interchangeably.
The number of faces in a lattice of side-length $L$ is $\binom{4}{2}L^4 = 6L^4$.
This follows from the fact that every face is uniquely determined by a vertex $v$ and two coordinates $i,j \in \{x,y,z,w\}$, so that the face with base-point $v$ lies in the $i$-$j$-plane.
More generally, the number of $k$-dimensional objects in the lattice (1-dimensional objects would be edges or 3-dimensional objects would be cubes) is given by~$\binom{4}{k}L^4$.
There are stabilizer checks for every edge and cube in the lattice.
A stabilizer check associated with a particular edge acts as Pauli-$X$ on all faces which are incident to it (see \autoref{fig:4DTCstabilizer} (left)), whereas a stabilizer check associated with a particular cube acts as Pauli-$Z$ on all faces incident to this particular cube (see \autoref{fig:4DTCstabilizer} (right)).
All stabilizer checks act on 6 qubits and each qubit is acted upon by 8 stabilizer checks.

Consider an operator $E$ acting as Pauli-$Z$ on some subset of qubits.
For $E$ to commute with an $X$-check their overlap has to be even.
Hence, to commute with all $X$-checks $E$ can not contain any connected components that have a boundary where a single face is incident to an edge.
Geometrically this means that~$E$ has to be a collection of boundaryless (closed) surfaces.
If $E$ itself is the boundary of a 3D volume inside the hypercubic lattice then it is the product of all $Z$-stabilizer elements inside this volume.
But there are other examples of surfaces which do not have a boundary:
Since the topology of the lattice is non-trivial we can consider a sheet extending over the whole $x$-$y$-plane.
Due to the periodicity of the lattice this sheet has no boundary, but is not itself the boundary of any 3D volume and thus it is not the product of any $Z$-stabilizers.
It is therefore a logical operator.
There is one such operator for every plane in the lattice.
Each plane is labelled by two coordinates ($x$-$y$-plane, $x$-$z$-plane, \dots) so that the 4D toric code encodes $\binom{4}{2} = 6$ logical qubits.
A sheet extending through the whole lattice consists of at least $L^2$ faces which means that the 4D toric code has distance growing quadratically with the number of physical qubits.
The parameters of the code are $[[n=6L^4, k = 6, d = L^2]]$.

In \cite{TNthreshold4D} the authors argue that the threshold of the 4D toric code against independent bit-flip and phase-flip errors is approximately $p_c \approx 0.11$ under minimum-weight decoding.
This is the same threshold as for the 2D toric code and the surface code.

\subsection{Syndromes}
A feature of the 4D toric code is that the set of violated stabilizer checks form extended objects:
since we pick up the boundary of an error (which is a collection of surfaces), the syndrome will always form closed loops as opposed to a set of points for the 2D toric code or surface code.
A minimum-weight decoding, similar to minimum-weight perfect matching for the 2D toric code, is finding a minimum-area surface among all surfaces that have the syndrome as its boundary.
It is known that the problem of finding such a minimum-are surface can be solved efficiently when the lattice is three-dimensional \cite{Sminimumareasurface}, but it is open whether it can be solved efficiently for a 4D lattice.
However, there are several decoders with worse error-correction performance than the minimum weight decoder but which are computationally efficient \cite{DKLP,Hdecoding,Pthesis,BDDTlocaldecoders,Atoomphd}.
The common way these decoders operate is by iteratively shortening the syndrome loops by flipping nearby faces.

\begin{figure}[h]
\begin{center}
\begin{tikzpicture}
\pgfmathsetmacro{\L}{1}
\draw[ thick, black,fill=gray, fill opacity=0.2] (0,0,0) -- ++(-\L,0,0) -- ++(0,0,-\L) -- ++(\L,0,0) -- cycle;
\draw[ thick, black,fill=gray, fill opacity=0.2] (\L,0,0) -- ++(-\L,0,0) -- ++(0,0,-\L) -- ++(\L,0,0) -- cycle;

\draw[ thick, black,fill=gray, fill opacity=0.2] (0,0,\L) -- ++(-\L,0,0) -- ++(0,0,-\L) -- ++(\L,0,0) -- cycle;
\draw[ thick, black,fill=gray, fill opacity=0.2] (\L,0,\L) -- ++(-\L,0,0) -- ++(0,0,-\L) -- ++(\L,0,0) -- cycle;

\draw[ thick, black,fill=gray, fill opacity=0.2] (0,0,0) -- ++(0,-\L,0)  -- ++(0,0,-\L) -- ++(0,\L,0) -- ++(0,0,\L);
\draw[ thick, black,fill=gray, fill opacity=0.2] (0,0,0) -- ++(0,\L,0) -- ++(0,0,-\L) -- ++(0,-\L,0);

\draw[ thick, black,fill=gray, fill opacity=0.2] (0,0,\L) -- ++(0,-\L,0)  -- ++(0,0,-\L) -- ++(0,\L,0) -- ++(0,0,\L);
\draw[ thick, black,fill=gray, fill opacity=0.2] (0,0,\L) -- ++(0,\L,0) -- ++(0,0,-\L) -- ++(0,-\L,0);

\draw[ thick, black,fill=gray, fill opacity=0.2] (0,0,0) -- ++(-\L,0,0) -- ++(0,-\L,0) -- ++(\L,0,0) -- cycle;
\draw[ thick, black,fill=gray, fill opacity=0.2] (\L,0,0) -- ++(-\L,0,0) -- ++(0,-\L,0) -- ++(\L,0,0) -- cycle;
\draw[ thick, black,fill=gray, fill opacity=0.2] (0,\L,0) -- ++(-\L,0,0) -- ++(0,-\L,0) -- ++(\L,0,0) -- cycle;
\draw[ thick, black,fill=gray, fill opacity=0.2] (\L,\L,0) -- ++(-\L,0,0) -- ++(0,-\L,0) -- ++(\L,0,0) -- cycle;

\draw[ thick, red] (-\L,0,0) -- ++(2*\L,0,0);
\draw[ thick, red] (0,-\L,0) -- ++(0,2*\L,0);
\draw[ thick, red] (0,0,-\L) -- ++(0,0,2*\L);
\draw[ thick, black,fill=red,] (0,0) circle (0.1);
\end{tikzpicture}
\end{center}
\caption{The dependency of the edge stabilizers which are incident to a common vertex. Taking the product of all edge-stabilizers (red) incident to a common vertex (red) gives the identity.}
\label{fig:loclindep}
\end{figure}
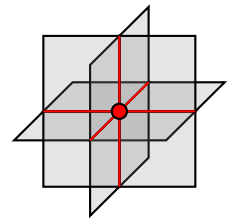

The fact that the syndromes form closed loops can be understood in terms of local linear dependencies between the stabilizer elements:
Taking all $X$-checks (edges) which are incident to a particular vertex gives the identity operator.
This is the case because for every triple of vertex, edge and face which are pairwise incident to one another we can always find one more edge which is incident to the same vertex and face (see \autoref{fig:loclindep}).
Similarly, taking all $Z$-checks (cubes) of one hypercube gives the identity since there are two cubes are overlapping on every face of the hypercube.

Each of those local linear dependencies can be interpreted as a (classical) parity check on the syndrome.
More explicitly: Both the $X$-syndrome and the $Z$-syndrome are encoded by a classical linear code.
The code words of this code are the {\em valid syndromes}.
What are the parameters of this code? - There is a bit for every $X$-check (edge) in the lattice, so the block size of the code is $4L^4$.
There is a local dependency for every vertex.
However, the local dependencies themselves are not independent.
Taking the product over all vertices and all edge-checks incident to this vertex gives the identity since every edge is incident to two vertices.
Hence, the number of independent checks in the classical linear code encoding the syndrome information is $L^4 - 1$.
The encoded syndrome information therefore contains $4L^4-(L^4-1) = 3L^4 + 1$ bits.
The distance of the classical code is 4 since adding the boundary of a face takes us from one valid syndrome to another valid syndrome.

Since the syndrome of the 4D toric code is encoded it has some build-in robustness against syndrome errors.
In comparison, it is known that the 2D toric code does not have a decoding threshold in the presence of syndrome noise.
The parity measurements have to be repeated and the record of the repeated syndrome measurements is decoded.
This essentially implements a repetition code in time.
Repeating measurements is not necessary for the 4D toric code to have increased error suppression capabilities with larger system size.
The fact that measurements are not repeated is referred to as {\em single-shot measurements}.
It has been shown analytically in \cite{bombin2015single} that a single-shot decoder can have a threshold and single-shot decoders for the 4D toric code have analyzed numerically in \cite{BDDTlocaldecoders,4drenormalization,arakawa2005self}.

As 4D space is hard to visualize it is useful to consider the {\em 3D toric code} to gain some geometric intuition.
It is defined on a 3D cubic lattice with periodic boundaries. 
Similarly to the 4D toric code the qubits are identified with the faces of the lattice and the $X$-stabilizer checks act on all faces incident to an edge (weight 4) while the $Z$-stabilizer checks act on all faces incident to a cube (weight 6).
A boundaryless sheet acting as Pauli-$Z$ commutes with all $X$-stabilizer checks just as for the 4D toric code.
The $X$-stabilizer checks also satisfy the local linear dependencies shown in \autoref{fig:loclindep}.
To commute with all $Z$-stabilizer checks it suffices to take a closed loop of Pauli-$X$ operators in the dual lattice.
The existence of these string-like logical operators results in a threshold which is generally lower under minimum-weight decoding \cite{TNthreshold4D,wang2003confinement}.

%% file: machine_learning.tex
To have a better understanding of our decoding procedure, it is useful to first give a small overview of machine learning.
This will provide insights to advantages and limitations of our procedure, and possible ways to improve it.
For more extensive review on machine learning and in particular neural networks, we refer to \cite{Mmachinelearning,Nmachinelearning,Gmachinelearning}.

Machine learning is often categorized into supervised learning, unsupervised learning and reinforcement learning, each has a different focus.
However, all of them usually involve the use of some models.
In this paper, the word ``model'' simply means a family of functions $f_\alpha$ or subroutines $s_\alpha$, where the (most likely multidimensional) parameter $\alpha$ needs to be trained by certain learning algorithms.
Below we give a not very comprehensive introduction to these three kinds of learning
\begin{itemize}
	\item {\em Supervised learning} is to find $\alpha$ such that $f_\alpha =  f_\text{hidden}$, where $f_\text{hidden}$ is given implicitly by a dataset of input-output pairs $(x_i,y_i=f_\text{hidden}(x_i))$.
	\item {\em Unsupervised learning} refers to many different tasks which are related to finding structures in a dataset. In contrary to supervised learning, there is no desired output for each entry in the dataset. For example, considering tasks in error correction, inferring properties of the error model from a dataset of syndromes or using the dataset to generate new similar syndromes can be viewed as unsupervised learning, as well as some tasks studied in~\cite{combes2014situ}.
	\item {\em Reinforcement learning}
    is concerned with how agents ought to take actions in an environment so as to maximize some notion of cumulative reward.
    However, in this paper we will use the word ``reinforcement learning'' to denote the optimization of a subroutine $s_\alpha$ of a program, so that at the end of the program a predefined score is maximized.
\end{itemize}
For complicated tasks, it is beneficial to combine these approaches. A notable example which highlights this is the recent success of the AI  \textit{AlphaGo}~\cite{silver2016mastering,silver2017mastering}.
We give a short summary of this combined approach in \aref{appendix:alphago} and discuss some ideas for applying it to the decoding problem of quantum codes.

One major difficulty encountered in machine learning is the overwhelmingly large input/state space.
Here we will only give a short explanation of how this difficulty arises in supervised learning.
Recall that we are given a dataset $(x_i,y_i)$ with the goal being approximating $f_\text{hidden}$.
If the input space is large, then for a random $x$, it is unlikely that $x$ is ``similar'' to any of the $x_i$ from the dataset.
Therefore, it is hard to guess the value of $f_\text{hidden}(x)$ and approximate it by $f_{\alpha}(x)$.
This is why often the most important part of a machine learning project is to choose a suitable model, usually by encoding the prior knowledge of the task into it.
The model we use in this paper is neural networks which we introduce in the following section.

\subsection{Basics of Neural Networks}

\label{sec:neural_network_basics}

A {\em neural network} is a directed, multipartite graph consisting of {\em layers} $l = 0,\dotsc,L$.
The vertices in layer $l$ are connected to vertices in the following layer $l+1$.

\begin{figure}[h]
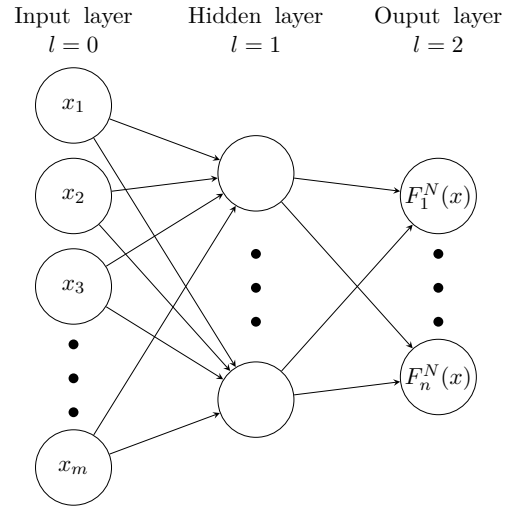

	\begin{center}
		\includestandalone[scale=0.8,mode=buildnew]{figures/neural_network_schematic}
	\end{center}
	\caption{A neural network consisting of $3$ layers. The network takes input $x \in \{0,1\}^{m}$ which is represented as the first layer of neurons. The values of neurons in the hidden layer and the output layer are given by the function $f_{w,b}:\real^q \rightarrow [0,1]$ evaluated on their input (indicated by $q$ incoming arrows). The parameters $w \in \real^{q}$ and $b\in \real$, called weights and bias, can be different for each neuron. The values of the neurons in the last layer are the output of the network $F^N(x) \in \real^{n}$.}
	\label{fig:nnschematic}
\end{figure}

The vertices of the network are called {\em neurons}.
The main idea behind the neural network is the following:
Each neuron  computes a primitive non-linear function $f_{w,b}:\real^q \rightarrow\real$, for which the input values are given by neurons of the previous layer connected to it.
The subscripts $w \in \real^{q}$ and $b\in \real$ are parameters which can differ for each neuron.
Before we discuss the function $f_{w,b}$ in more detail let us first understand how the network performs a computation.
The neurons in the first layer do not have any predecessors and their output is simply set to be the input of the network, which is a bit string $x \in \{0,1\}^{m}$.
The values of the neurons in the last layer are interpreted as the output of the network.
We see that the network describes a function $F : \{0,1\}^{m} \rightarrow [0,1]^{n}$.
The first layer $l=0$ is called the {\em input layer} and the last layer $l=L$ is called the {\em output layer}.
All other layers $l = 1,\dotsc,L-1$ are called {\em hidden layers} since they are considered to be internal to the network.

\begin{figure}[h]
	\begin{center}
		\begin{tikzpicture} 
		\begin{axis}[xmin=-5,xmax=5,no markers,samples=50,height=6.5cm, width=6.5cm,grid=both]
		\addplot {1/(1+exp(-x))} node[below left]{$\sigma(z)$};
		\addplot[domain=-5:0] {0} node[right]{$\Theta(z)$};
		\addplot[dashed]  coordinates {(0,0)(0,1)};
		\addplot[domain=0:5] {1};
		\end{axis}
		\end{tikzpicture}
	\end{center}
	\caption{The sigmoid function $\sigma$ is a smooth version of the Heaviside step function $\Theta$.}
	\label{fig:sigmoid}
\end{figure}
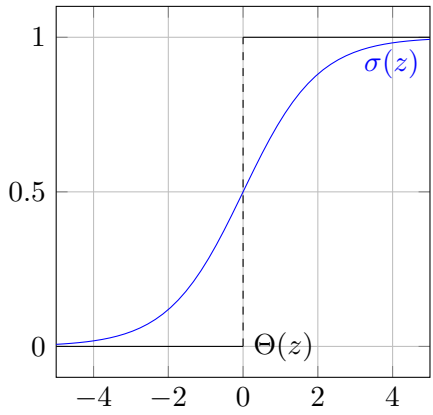

The parameters $w$ and $b$ are called {\em weights} and {\em biases}.
They define a linear map $w\cdot y + b$ where $y$ is the input of the neuron.
The function that each neuron computes has the form:
\begin{align}
f_{w,b}:\real^q \rightarrow [0,1],\quad y \mapsto \sigma(w\cdot y + b)
\end{align}
where in this paper $\sigma$ is either $\tanh$ or the {\em sigmoid function}
\begin{align}
\sigma(z) = \frac{1}{1+\exp(-z)}
\end{align}
which is plotted in \autoref{fig:sigmoid}.
Both non-linear functions can be thought of as a smoothed step function.
The smoothness of $\sigma$ is important to the training of the neural networks as it is based on optimizing certain objective function with gradient descent.

Let us have a look at a small example which gives some intuition why neural networks are able to perform interesting computations:
Consider a single neuron which takes $q = 2$ inputs and has weights $w = (-12,-12)$ and bias $b = 17$.
The neuron computes the following values for each input:
\begin{align}
\begin{split}
f(0,0) &= \sigma(17) \approx 1\\
f(1,0) &= \sigma(5) \approx 1\\
f(0,1) &= \sigma(5) \approx 1\\
f(1,1) &= \sigma(-7) \approx 0.
\end{split}
\end{align}
We observe that these are approximately the input/output relations of the NAND gate.
The approximation can be made arbitrarily close by increasing the absolute values of the weights $w$ and the bias $b$.
Since any Boolean function $F : \{0,1\}^m \rightarrow \{0,1\}^n$ can be computed by a network of NAND gates, there consequently also exists a representation of $F$ as a neural network.
However, in practice it is more efficient to adapt the connectivity of the network to the problem at hand.

The process of {\em training} the network, i.e. gradually adapting the values of the weights $w$ and biases $b$ to make the network a desired function, is described in \aref{appendix:training_neural_network}.

\subsection{Convolutional Neural Networks}
\label{sec:CNN}
Compared to fully connected neural networks, convolutional neural networks (CNN) require much fewer parameters to describe.
We will start with the definition of one convolutional layer.
The input to the layer resides on a $D$-dimensional lattice of size $L^D$.
On each lattice site, there is a $d$-dimensional vector $x_\mathbf{u} \in \real^d$, where the subscript $\mathbf{u} \in \mathbb{Z}_L^D$ (we use $\mathbb{Z}_L$ to denote integer in range $[0,L-1]$).
We define the kernel to be a vector $K_{\mathbf{u},i}$, where $\mathbf{u} \in \mathbb{Z}_n^D$ and $i \in \mathbb{Z}_d$. With a slight abuse of notation, we will say such a kernel has size $n^D$. The convolution is then
\begin{align}
\label{eq:convolution}
y_\mathbf{v} = \sum_{\mathbf{u} \in \mathbb{Z}_n^D} \sum_{i \in \mathbb{Z}_d} x_{v-u, i} K_{u, i},
\end{align}
where $x_{v-u,i}$ is the $i$th element of $x_{v-u}$, and each element of $\mathbf{v}$ has range $[n-1, L-1]$.
The index $v$ of the output $y$ can be shifted according to needs (e.g. to $\mathbb{Z}_{L-n+1}^D$).
In this paper, we solely work with input that has the periodic boundary condition.
Thus, we can indeed treat the indices $\mathbf{u}$ of input $x_\mathbf{u}$ as elements of the cyclic group $\mathbb{Z}_L^D$.
The convolution given in \autoref{eq:convolution} can then be modified accordingly such that $\mathbf{v} \in \mathbb{Z}_L^D$.
If there are $r$ different kernels, we can apply \autoref{eq:convolution} for each kernel individually and obtain $y_{\mathbf{v},i}$, where $i \in \mathbb{Z}_r$.
We will say the resulted output $y_{\mathbf{v},i}$ has $r$ channels, and further convolution can be again applied on $y_{\mathbf{v},i}$ according to \autoref{eq:convolution}.
Non-linear functions are usually applied after each convolution.
In this paper, convolutional neural networks are neural networks with only convolutional layers (where in computer vision, the convolutional neural networks usually contains coarse-grain layers, etc).

The connectivity of a 1D convolutional neural networks is illustrated in \autoref{fig:1d_conv}.
We can see that a convolutional layer is simply a neural network layer with local connectivity and the weights are translationally invariant.

%% file: figures/neural_network_schematic.tex
\tikzset{%
    every neuron/.style={
        circle,
        draw,
        minimum size=1.25cm
    },
    neuron missing/.style={
        draw=none, 
        scale=4,
        text height=0.333cm,
        execute at begin node=\color{black}$\vdots$
    },
}

\begin{tikzpicture}[x=1.5cm, y=1.5cm, >=stealth]

\foreach \m/\l [count=\y] in {1,2,3,missing,4}
\node [every neuron/.try, neuron \m/.try] (input-\m) at (0,2.5-\y) {};

\foreach \m [count=\y] in {1,missing,2}
\node [every neuron/.try, neuron \m/.try ] (hidden-\m) at (2,2-\y*1.25) {};

\foreach \m [count=\y] in {1,missing,2}
\node [every neuron/.try, neuron \m/.try ] (output-\m) at (4,1.5-\y) {};

\foreach \l [count=\i] in {1,2,3,m}
\draw (input-\i) node {$x_{\l}$};


\foreach \l [count=\i] in {1,n}
\draw (output-\i) node {$F^N_\l (x)$};

\foreach \i in {1,...,4}
\foreach \j in {1,...,2}
\draw [->] (input-\i) -- (hidden-\j);

\foreach \i in {1,...,2}
\foreach \j in {1,...,2}
\draw [->] (hidden-\i) -- (output-\j);

\foreach \l [count=\x from 0] in {Input, Hidden, Ouput}
\node [align=center, above] at (\x*2,2) {\l $\:$ layer\\ $l = \x$};

\end{tikzpicture}

%% file: machine_learning_for_decoding.tex
Decoding quantum error correcting codes can be considered both as a supervised and reinforcement learning problem.
\begin{description}
	\item [Supervised learning:] In most scenarios, we can generate pairs of (syndrome, correct decoding), either by simulation on classical computer or data from real experiments. Thus, we can use the pairs to train suitable models.
	\item [Reinforcement learning:] Training certain types of decoders is more naturally considered as reinforcement learning problem. For example, we can leave some freedom in the cellular automaton decoder to be trained, with the goal of optimizing the memory time. In this setting, there is not a clear correct answer at each time step, and therefore we cannot directly train the cellular automaton with a input-output relation.
\end{description}
One major difficulty of applying machine learning to decoding is the variable lattice size (or in general, variable code size of some code family).
Note that usually learning algorithms run on one code size at a time, and it is fair to assume that the training time will grow when we increase the code size, possibly exponentially. Thus, it is not trivial to train a machine learning decoder on a larger lattice to utilize the error suppression ability provided by topological codes.
Moreover, if we restrict the machine learning models to neural networks, it is tempting to directly train a fully-connected neural nets with the pair (syndrome, correct decoding). While this is theoretically possible, in practice it is very hard to achieve for large lattices, due to the following:
\begin{enumerate}
	\item Decoding quantum error correcting codes is a quite complicated task. In order for the fully-connected neural nets to approximate the input-output relation, they need to have large amount of parameters (either by having many layers or many neurons in each layer). This might already require too many computational resources to set up the nets or train them.
	\item A neural net with large amount of parameters in turn requires large amount of training data, in order to not overfit.
\end{enumerate}
See \aref{appendix:nearest_neighbor} for a relevant discussion on obstacle of applying a baseline machine learning algorithm to decoding problem on large lattice.

\subsection{Architecture and Training}
\label{subsection:architecture_and_training}
In this subsection, we will only concern about Pauli-$X$ stabilizer checks.
We can pack the $6L^4$ qubits into a multi-dimensional array of shape $(L,L,L,L,6)$, and the Pauli-$X$ stabilizer checks into one of shape $(L,L,L,L,4)$.
We choose to use the following architecture for decoding:
\begin{description}
	\item [Architecture:] Given an $X$-syndrome with shape $(L,L,L,L,4)$, a convolutional neural net (CNN) is applied, with an output of shape $(L,L,L,L,6)$. Pauli-$X$ is applied to one or multiple qubits according to the output, and the error syndrome is updated accordingly. This process is repeated until no syndrome is left or certain condition for halting is met. For the particular training process we will describe shortly after, we choose to flip qubits corresponding to the largest values in the outputs from the CNN.
	\item [Training:] It is not clear how the CNN should decide which qubits to flip, thus a natural approach would be optimizing the CNN to achieve the best memory time. However, we decide to explicitly train the CNN to estimate whether a bit-flip has occurred on each qubit. While not optimal, it allows a much faster training process and it is less prone to finite size effect potentially caused by reinforcement learning on a small lattice.
\end{description}
This architecture has a few desired properties. First, the convolutional neural net is a constant-depth (with respect to $L$) and translational-invariant local circuit. As a result, each element of the $(L,L,L,L,6)$ output is determined by its surrounding region in the same way. Note that this also means that the training only needs to be done on a constant size region, which is a huge advantage for high dimensional lattices as the computational resource needed grows very fast with respect to $L$. Once we trained the CNN on a lattice with size $L$, we can naturally extend it to a larger size $L' >L$, therefore we can evaluate the performance of a fixed CNN on several different size $L$.  This is crucial for discussing the error suppression ability by increasing the system size, or performing an estimation of the decoding threshold.
We shall point out that the architecture together with training procedure is designed with 4D toric code in mind, where multiple local decoders have already shown a threshold behavior.

To some degree, the aim of our neural network decoder is to develop a local decoding strategy such as those analyzed in~\cite{BDDTlocaldecoders}.
In order to see this, we will give a brief introduction to the DKLP rule and the Toom's rule.
For the DKLP rule, a qubit is flipped if 3 or 4 adjacent edge stabilizer checks are violated, and is flipped with probability $1/2$ if only two are violated.
This update rule should not apply simultaneously to faces that share an edge.
Toom's rule is defined for a 2D lattice, where each face is associated with a degree of freedom which can take the values $+1$ and $-1$. In each step of the Toom's rule, the value of a face will be flipped if it is different from both its `north' and `east' faces' values.
It can be applied to the 4D toric code by applying it to every 2D plane of the 4D hypercubic lattice (e.g. applying to all $x$-$y$-planes and then applying to all $x$-$z$-planes, etc).
It is not hard to see that for both rules, the criteria which decides whether a qubit should be flipped or not can be written as a neural network.
After all, the neural networks can approximate any function on a finite input spaces.
Intuitively, the DKLP rule tries to flip qubits that likely have an error on them, which our neural network decoder also intends to achieve due to the specific training procedure.
On the other hand, Toom's rule is not designed with this goal in mind.
Therefore, it will not be obtained by our training procedure.
This serves as a reminder that some good decoders require a different training procedure to find, e.g. some kind of reinforcement learning.
It is worth noting that the schedule components of these rules (i.e. applying Toom's rule to 2D planes sequentially and applying DKLP rule to non-adjacent faces) are not contained in the architecture of our decoder.
If the goal is to accurately simulate the DKLP and Toom's rules, we can introduce a clock variable as an additional input on each face.

One detail of our architecture is that, as we mentioned in the architecture paragraph, we flip qubits corresponding to the largest values in the outputs from the CNN. In order to reduce the computational time of decoding, we can flip multiple qubits based on a single evaluation of the CNN. In the numerical simulation described in \autoref{sec:Num}, the number of qubits to be flipped is chosen to be $\# (\text{current violated syndrome checks})/x$, where $x$ is around 50.
We did not study the effect of different choice of $x$.
It can be viewed as a tunable parameter that requires further optimization.
Another reasonable way is to flip the qubits whose corresponding values from the CNN's output are larger than a pre-determined threshold.
If this approach is used, the decoder will be completely local with respect to the 4D lattice.

We also add an additional step in the architecture when decoding syndromes from perfect measurements of check operators, which is to apply the ``parallel line decoder'' introduced in \aref{sec_parallel_line_decoder}.
The motivation comes from the observation that without the parallel line decoder, a large percentage of the failures are resulted from the decoder getting stuck at syndromes that only contains a few pairs of parallel lines (e.g. \autoref{fig:parallel_line_decoder}).
Failures of this type are called ``energy-barrier limited decoding'' in~\cite{BDDTlocaldecoders}.
As we do not focus on building a completely local decoder, we choose to use this simple and fast subroutine to improve the performance.

%% file: numerical_analysis.tex
\subsection{Error Model and Setup}
\label{sec_error_model_and_setup}

We perform a numerical analysis of the neural network decoder by Monte Carlo simulations.
We consider two error models:
(a) In the  first error model we assume that the measurements of the check operators can be done perfectly.
We consider the independent $X$-$Z$-error model in which a Pauli-$X$ and a Pauli-$Z$ are applied independently to each qubit with probability $p$.
(b) In the second error model we take measurement errors into account.
The errors on the qubits are modeled as in (a).
Additionally each outcome of a syndrome measurement is flipped with probability~$q$.
The error correction procedure of the Pauli-$X$ errors and the Pauli-$Z$ errors can be done independently as each parity-check can only detect either type and all parity-checks are pairwise commuting.

For error model (a) where measurements are perfect we numerically estimate the {\em rate of logical errors~$\plog$} depending on the physical error rate~$p$.
The rate of logical errors $\plog$ is the probability of the neural network decoder to apply a non-trivial logical operator (or exceeding a time limit).
We can estimate $\plog$ for a fixed $p$ as follows:
First, we sample an error~$E$ by applying a Pauli-$X$ or Pauli-$Z$ each with probability~$p$.
We then compute the result of the syndrome check measurements $s$ and give it to the neural network decoder which determines a recovery operator~$R$.
After the application of~$R$ we are back in a code state.
The decoder was successful if the application of~$E$ and~$R$ acts trivially on the code state.
This can be checked efficiently since any operator which leaves the code space as a whole invariant acts non-trivially within the code space if and only if it commutes with all other logical operators.

For error model (b), which takes syndrome errors into account, the neural network decoder can in general not correct back to a code state since the decoder  has access to the erroneous measurement result only.
To obtain a threshold we estimate the {\em memory time}~$T$ which gives the average number of error correction rounds until a logical failure occurs.
If the physical error rate is below threshold the memory time will diverge with increasing lattice size $L$.
More concretely, in each error correction round, we first sample an error~$E$ by applying a Pauli-$X$ or Pauli-$Z$ each with probability $p$ and update the (noiseless) syndrome check measurement $s$ from the last round.
We then flip each bit in $s$ with probability $q$ to obtain the faulty syndrome check measurement $s_{\text{in}}'$, and feed $s_{\text{in}}'$ to the neural network decoder.
The decoder will output a list of position where Pauli-$X$ or Pauli-$Z$ are applied, so we can keep track of the noiseless syndrome measurement $s_{\text{out}}$ afterwards.
To evaluate whether a logical failure has occurred, we input $s_{\text{out}}$ to the decoder we use in the error model (a).
Note that we do not have to repeat the syndrome measurements as for the 2D toric code.

The networks we consider consist of a input layer which receives the result of the parity check measurements, one convolutional layer with kernel size $3^4$ (i.e. $n=3$ in \autoref{eq:convolution}) and then two convolutional layer with kernel size $1^4$.
The number of channels in the hidden layer is 15 for the noiseless syndrome measurement, and 20 for the noisy case.
The choices of these numbers here are mostly aimed for a large neural network without getting too slow or require too much memory to train.
After the linear map of the final convolutional layer, a softmax function is applied on all inputs:
\begin{align}
x_i \rightarrow \frac{e^{x_i}}{\sum_i e^{x_i}}
\end{align}
It is a widely used in the scenario when we want the output to be a probability distribution (i.e. non-negative numbers that sum to 1).
However, there is no obvious reason that the softmax layer is needed in our neural networks, as during the decoding process we only care about the relative order of the output numbers, which the softmax layer preserves.
The rational behind this is that among the limited number of neural networks we trained, the ones with the softmax layer performs slightly better.
Note that the softmax layer breaks the locality of the networks.
Nevertheless, if preferred, we can view softmax together with the cross-entropy as the modified cost function, while the neural networks remains local.

\begin{figure}
	\centering
	\includegraphics[width=\linewidth]{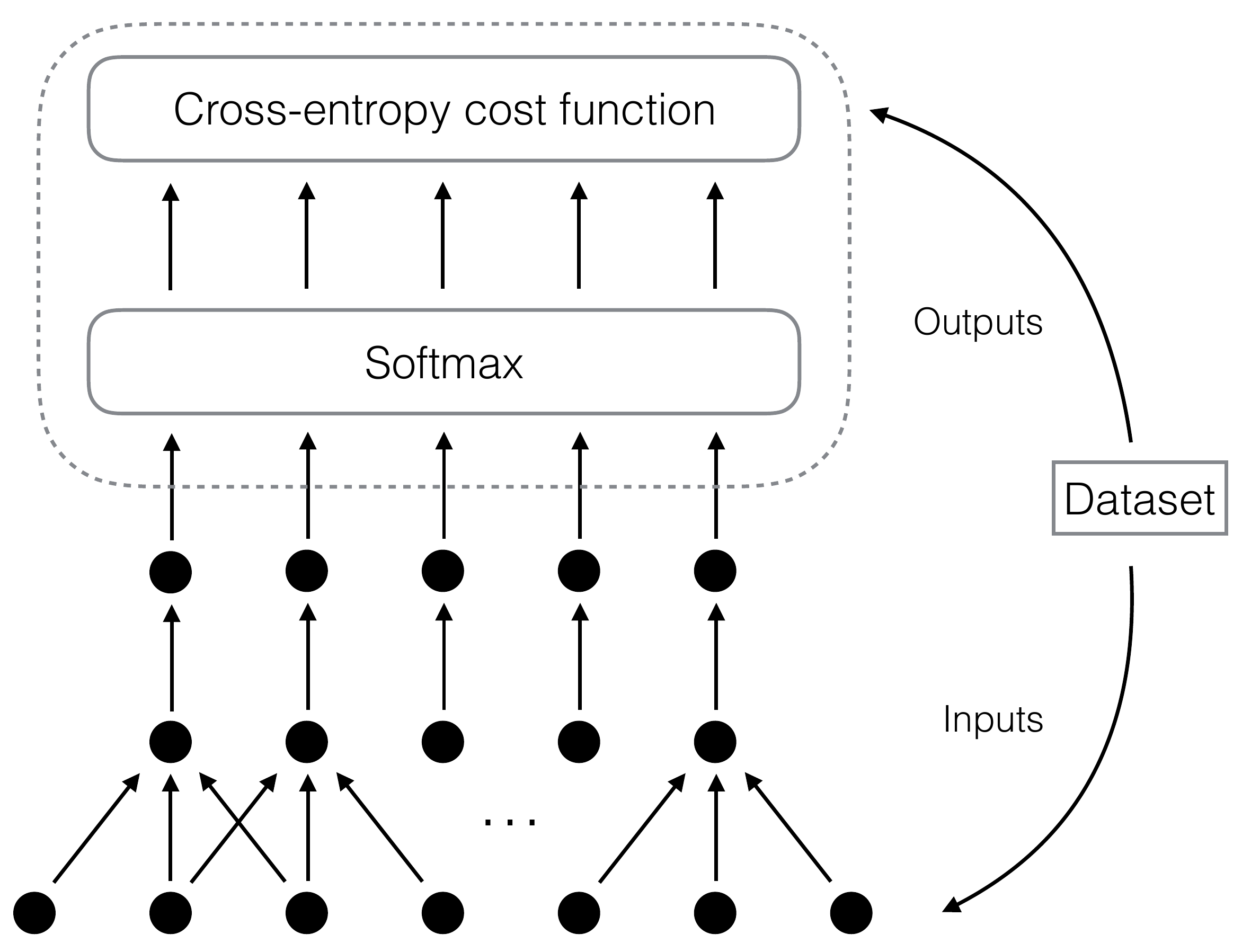}
	\caption{A 1D slice of the decoding neural network. The neurons in the first hidden layers depend on the neighbouring sites of the input layer, while each neuron in the later layers only depend on the neurons in the same sites in the previous layer.
		The input from the dataset is the error syndrome.
		The softmax layer does introduce non-locality in the network. However, we can effectively view softmax together with the cross-entropy as the modified cost function.}
	\label{fig_1d_slice_decoding_nn}
\end{figure}

A 1D slice of the neural network connectivity is shown in \autoref{fig_1d_slice_decoding_nn}.
It is obvious that without the softmax layer, each element in the output is determined only by a corresponding $3^4$ local region on the input.
The reason that we only have one convolutional layer with kernel size $3^4$ is mainly due to the faster runtime and likely will cause less finite-size effect on the lattice sizes we are testing on. 
See \aref{appendix:analysis_larger_network} for more discussion on using deeper neural networks.

 \makeatletter
 \tikzoption{canvas is xy plane at z}[]{%
   \def\tikz@plane@origin{\pgfpointxyz{0}{0}{#1}}%
   \def\tikz@plane@x{\pgfpointxyz{1}{0}{#1}}%
   \def\tikz@plane@y{\pgfpointxyz{0}{1}{#1}}%
   \tikz@canvas@is@plane
 }
 \makeatother

\begin{figure}
\begin{center}
\includestandalone[scale=1,mode=buildnew]{figures/network}
\end{center}
\caption{Illustration of the neural network. Each array of neurons is shown as a 2D square grid but in fact has the same dimensionality as the lattice (3D or 4D). The network consists of a single input layer (leftmost array) which receives the measurement results. The input layer is followed by three hidden layers. The number of channels in each hidden layer is 4 in this illustration. The final layer returns the probability distribution as output. A single convolution between the input layer and the first hidden layer is indicated in blue.}
\end{figure}

\begin{figure}
    \centering
    \begin{subfigure}[t]{0.45\textwidth}
        \centering
        \includegraphics[width=0.65\linewidth]{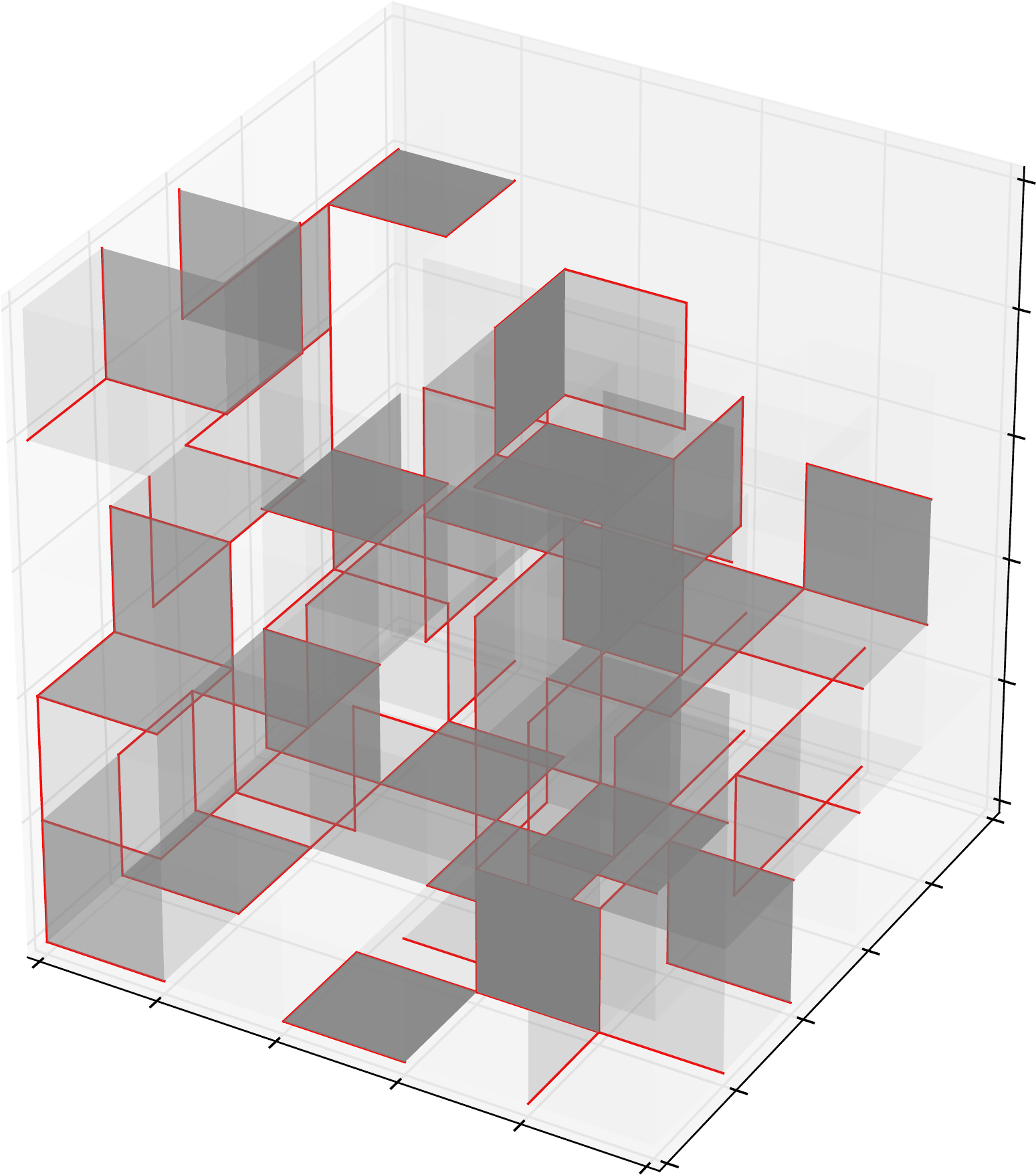}
        \caption{0 steps}
    \end{subfigure}
    \vspace{0.5cm}
    
    \begin{subfigure}[t]{0.45\textwidth}
        \centering
        \includegraphics[width=0.65\linewidth]{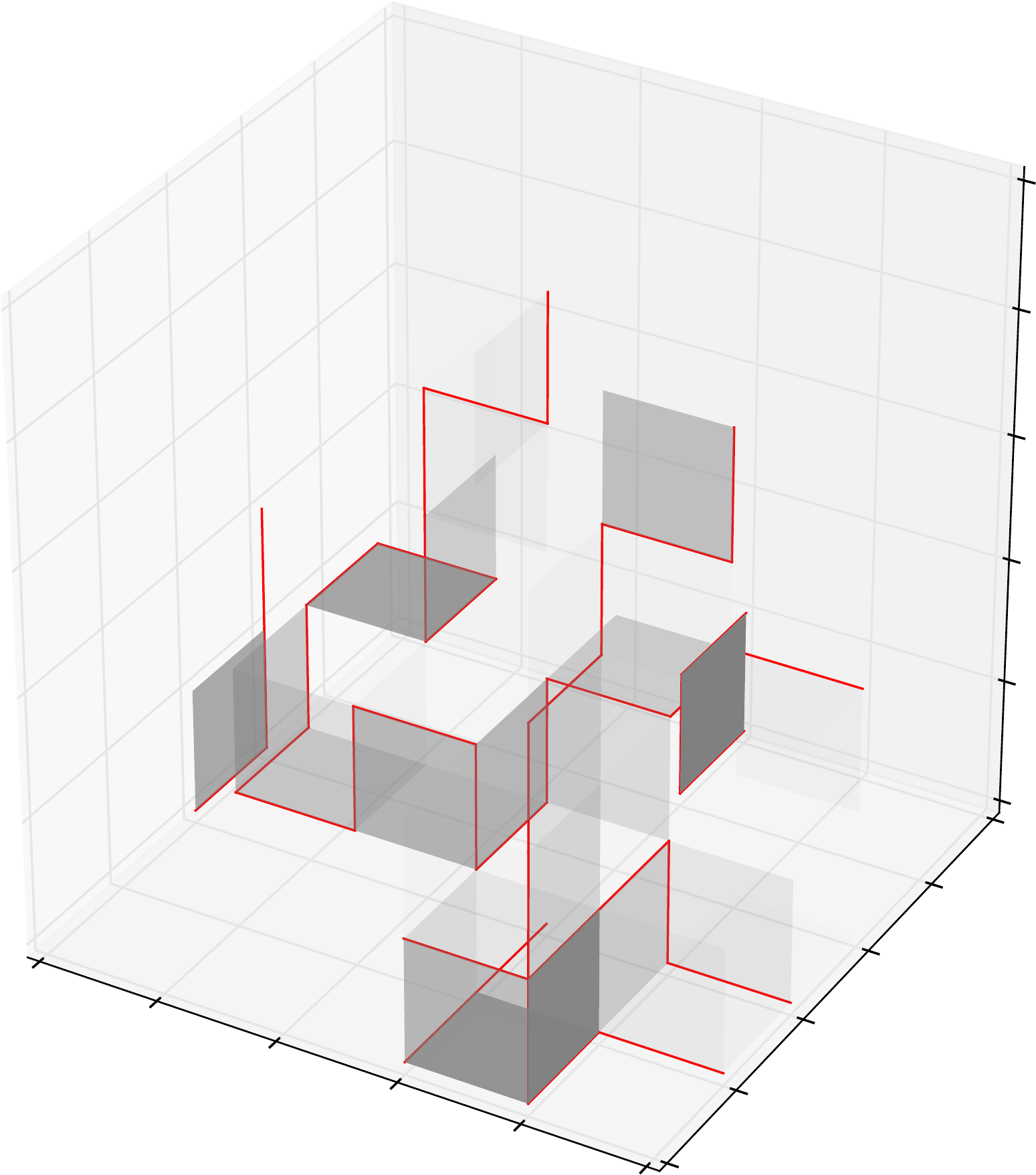}
        \caption{15 steps}
    \end{subfigure}
    \vspace{0.5cm}
    
    \begin{subfigure}[t]{0.45\textwidth}
        \centering
        \includegraphics[width=0.65\linewidth]{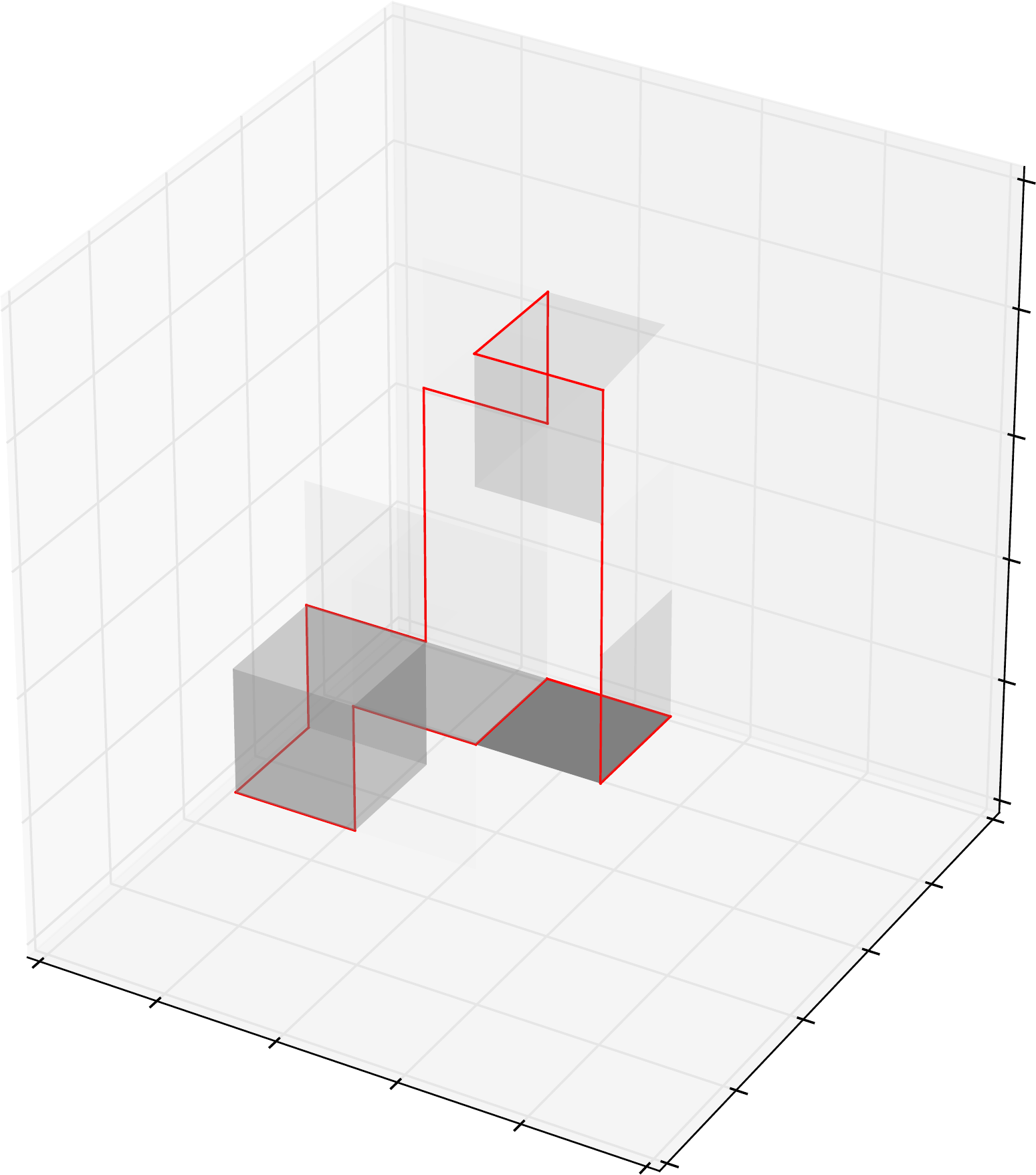}
        \caption{30 steps}
    \end{subfigure}
     
    \caption{Applying the neural network decoder to the 3D toric code with $L=5$. The syndrome is highlighted in red. The neural network outputs a probability distribution over the faces indicating where it believes an error to be present. The probability of each face is indicated by its opaqueness. Each figure shows the current syndrome and the output of the network during the decoding. In each step the decoder flips the face with the highest probability.}
    \label{fig:nndecoder3D}
\end{figure}

\subsection{Training}

We trained two neural networks for error model (a) and (b) described in \autoref{sec_error_model_and_setup} respectively.
However, the same neural network will be used for decoding syndromes generated according to different error probabilities.
Roughly speaking, the training of the neural networks is done using gradient descent (see \autoref{sec:neural_network_basics}). The inputs are error syndromes of shape $(L,L,L,L,4)$.
The outputs have shape $(L,L,L,L,6)$, where an element is equal to $1/N$ if an error happened on the corresponding qubit and equal to $0$ otherwise, with $N$ being the total number of errors.
The normalization is done to match the normalization done by the softmax layer.
We use a variation of the gradient descent algorithm called \textsc{Adam}~\cite{kingma2014adam}, which is included in the \textsc{Tensorflow} library.
The cost function we use is cross-entropy.
We also manually lower the learning rate of \textsc{Adam} when the decrease of cost function slows down.
For error model (a), we train the network with syndromes corresponding to error rate $p$ uniformly distributed from $3\%$ to $7\%$.
For error model (b), we train the network with error rate $p$ uniformly distributed from $2\%$ to $3\%$, and $q$ being a constant $2.5\%$.
These values were determined to be be the approximate locations of the thresholds in trial runs.

We also applied the neural network decoder to the 3D toric code under error model (a).
The convolutions chosen to be three dimensional in this case, but the structure of the neural network is identical otherwise.
The network is trained for $p$ at $17\%$ which is again just below the value of the threshold.

We do want to note that the details of the training are not very important (e.g. the variation of gradient descent we use, the concrete parameters we used for gradient descent, etc).
Indeed, the neural nets in this paper are fairly shallow compared to the neural networks used by the machine learning community at the present time.
Based on experience, any refined gradient optimizer should be able to train the networks decently well without fine-tuning the training parameters.
Additionally, we almost did not do any post-selection on the training of neural networks, and in general the decoder works reasonably well with a trained neural network.
We believe the performance of decoder observed in this paper is not a rare event.

\subsection{Performance}
\begin{figure}[htbp]
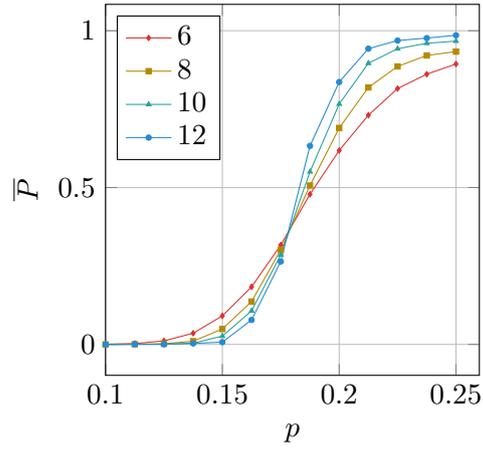
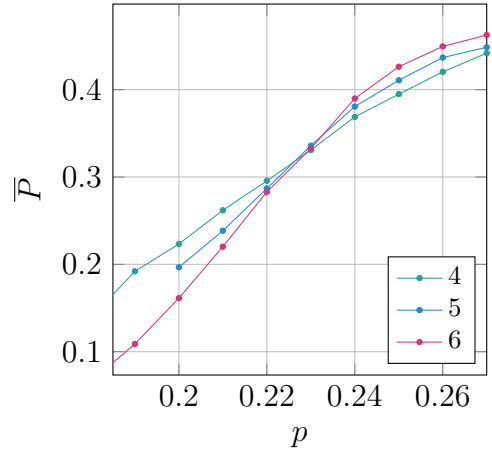

    \centering
    \begin{subfigure}[a]{0.45\textwidth}
        \includestandalone[scale=1,mode=buildnew]{figures/nn_noiseless3D}
        \caption{Neural network decoder}
    \end{subfigure}
    \vspace{0.5cm}

    \begin{subfigure}[a]{0.45\textwidth}
        \includestandalone[scale=1,mode=buildnew]{figures/failrate_global_3D}
        \caption{Minimum-weight decoder}
    \end{subfigure}
    \caption{(a) The results of the numerical simulation for the 3D toric code for $Z$-errors only, assuming perfect measurements. We considered system sizes $L=6,8,10,12$. The lines cross at around~$17.5\%$. (b)~Results for the minimum-weight decoder which has exponential run-time. The lines cross at around~$23 \%$. Note that the threshold of the line-like logical operator will be significantly lower (see \cite{TNthreshold4D}).}
    \label{fig:num3D}
\end{figure}

\begin{figure}[htbp]
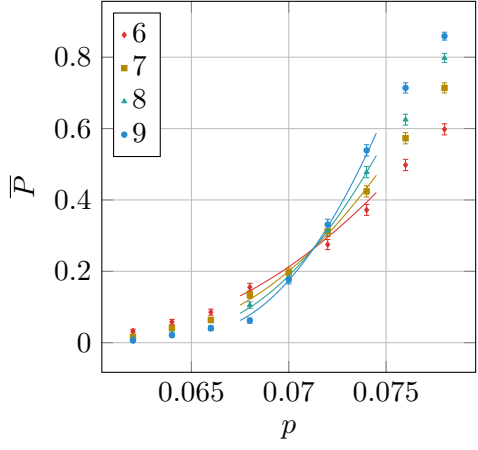
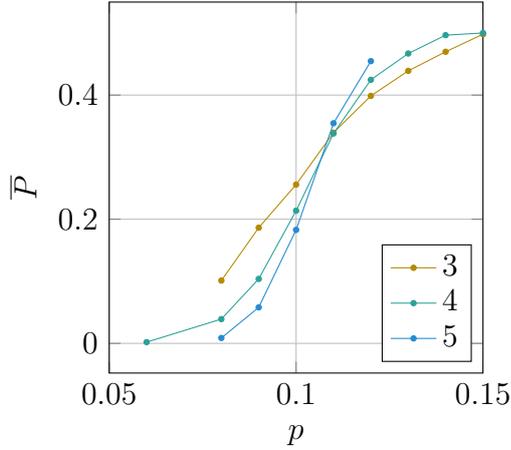

    \centering
    \begin{subfigure}[a]{0.45\textwidth}
        \includestandalone[scale=1,mode=buildnew]{figures/nn_noiseless4D}
        \caption{Neural network decoder}
    \end{subfigure}
    \vspace{0.5cm}
    
    \begin{subfigure}[a]{0.45\textwidth}
        \includestandalone[scale=1,mode=buildnew]{figures/failrate_global_4D}
        \caption{Minimum-weight decoder}
    \end{subfigure}
    
    \caption{(a) The results of the numerical simulation for the 4D toric code assuming perfect measurements. We considered system sizes $L=5,6,7,8$. The solid lines are given by the values of \autoref{eqn:scaling}. (b)~Numerical simulation for the minimum-weight decoder assuming perfect measurements. The lines cross at around $11\%$ which is in agreement with the numerical results of~\cite{arakawa2005self}.}
    \label{fig:num4D}
\end{figure}

To evaluate the performance, we follow the procedures described in \autoref{sec_error_model_and_setup}.
We will first discuss the results for error model (a) where we assume perfect stabilizer measurements.
As we mentioned in \autoref{subsection:architecture_and_training}, the parallel line decoder is applied after the neural network decoder when obtaining these results.
In \autoref{fig:num3D} and \autoref{fig:num4D}, we plot the logical error rates versus the physical error rates and compare them with the minimum-weight decoder.
The minimum-weight decoder is implemented by mapping the problem of finding a minimum surface with the syndrome as boundary to a linear integer programming problem.

We assume a scaling behavior in the variable 
\begin{align}
x = (p - p_c)L^{1/\nu}
\end{align}
to determine the critical error probability $p_c$.
We expand the logical error probability $\plog$ for small $x$ (around $p = p_c$ where the dependence on the system size $L$ is small):
\begin{align}\label{eqn:scaling}
 \plog(p,L) = A + Bx + Cx^2 
\end{align}
For the 4D toric code we obtain by fitting \autoref{eqn:scaling} to the data for $p = 0.066, 0.068, 0.072, 0.074$ (see \autoref{fig:num4D}).
The fitting parameters $p_c$ and $\nu$ were determined by a non-linear fit:
\begin{align}
    p_c = 0.071 \pm 0.003 ,\quad \nu = 0.65 \pm 0.02
\end{align}
This is comparable to the performance of the 4D renormalization group decoder described in \cite{4drenormalization} which achieves a threshold of $p_c=0.073\pm 0.001$ for the same error model.

\begin{figure}[htbp]
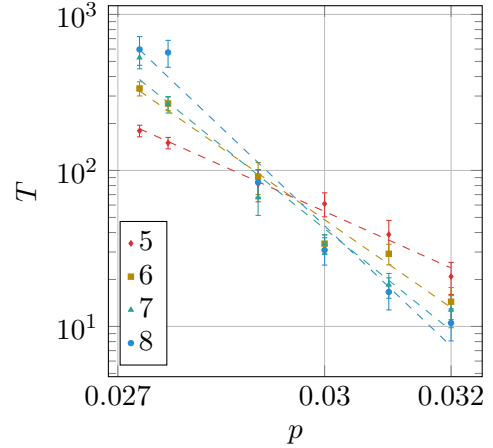

	\centering
	\includestandalone[scale=1,mode=buildnew]{figures/nn_noisy4D_fit}
	\caption{The results of the numerical simulation for the 4D toric code assuming noisy measurements. We considered system sizes $L=5,6,7,8$. The error bars show the $80\%$ confidence interval. The dashed lines indicate a linear fit in the $\log$-$\log$--plot. The memory time is expected to decay exponentially with~$p$ and diverge exponentially with increasing~$L$ for any fixed $p<p_c$. The data does not allow a precise estimation of the threshold~$p_c$. }
	\label{fig:memory_time}
\end{figure}

Below we will discuss error model (b) with measurement errors.
Here we will always set $p=q$, where $p$ is the error rate of the physical qubits, and $q$ is the error rate of each stabilizer parity measurement.
For clarification, we do not use the parallel line decoder for this error model.
In \autoref{fig:memory_time}, we plot the average memory time $T$ versus the error rate $p$.
Due to a time restriction on the computing cluster some runs were forced to halt where the simulated system was still in a correctable state.
In the worst case (large system size and low physical error rate) only around $10\%$ of the runs finished.
In order to obtain meaningful statistics from the data, we make the assumption that for a fixed $L$ and~$p$, the memory time follows an exponential distribution, and all unfinished runs have longer memory time than finished runs (which is called type II censoring on the right in the statistics literature).
Under these assumption, the mean (and its confidence interval) can be estimated according to~\cite{david2004order}.

%% file: figures/network.tex
\begin{tikzpicture}
 [x={(0.866cm,0.5cm)}, y={(-0.866cm,0.5cm)}, z={(0cm,1cm)}, scale=0.15]
\begin{scope}[canvas is xz plane at y=35]
	\draw[thick,draw=red] (5,5) rectangle (6,6);
\end{scope}  
 
\begin{scope}[canvas is xz plane at y=30]
	\draw [very thin, draw=gray, fill=gray, fill opacity=0.2] (3,3) grid  (11,11) rectangle (3,3);
	\draw[fill=red] (5,5) rectangle (6,6);
	\draw[fill=red] (5,6) rectangle (6,7);
	\draw[fill=red] (6,5) rectangle (7,6);
	\draw[fill=red] (6,6) rectangle (7,7);
	
	\draw[draw=black, fill=blue, fill opacity=0.2] (4,4) rectangle (8,8);
\end{scope} 

\begin{scope}[canvas is xz plane at y=20]
	\draw [very thin, draw=gray, fill=gray, fill opacity=0.2] (0,0) grid  (6,6) rectangle (0,0);
	\draw [very thin, draw=gray, fill=gray, fill opacity=0.2] (7,0) grid  (13,6) rectangle (7,0);
	\draw [very thin, draw=gray, fill=gray, fill opacity=0.2] (0,7) grid  (6,13) rectangle (0,7);
	\draw [very thin, draw=gray, fill=gray, fill opacity=0.2] (7,7) grid  (13,13) rectangle (7,7);
	
	\draw[draw=black, fill=blue, fill opacity=0.2] (1,1) rectangle (2,2);
	\draw[draw=black, fill=blue, fill opacity=0.2] (8,1) rectangle (9,2);
	\draw[draw=black, fill=blue, fill opacity=0.2] (1,8) rectangle (2,9);
	\draw[draw=black, fill=blue, fill opacity=0.2] (8,8) rectangle (9,9);
\end{scope}

\begin{scope}[canvas is xz plane at y=10]
	\draw [very thin, draw=gray, fill=gray, fill opacity=0.2] (0,0) grid  (6,6) rectangle (0,0);
	\draw [very thin, draw=gray, fill=gray, fill opacity=0.2] (7,0) grid  (13,6) rectangle (7,0);
	\draw [very thin, draw=gray, fill=gray, fill opacity=0.2] (0,7) grid  (6,13) rectangle (0,7);
	\draw [very thin, draw=gray, fill=gray, fill opacity=0.2] (7,7) grid  (13,13) rectangle (7,7);
\end{scope}
 
\begin{scope}[canvas is xz plane at y=0]
	\draw [very thin, draw=gray, fill=gray, fill opacity=0.2] (0,0) grid  (6,6) rectangle (0,0);
	\draw [very thin, draw=gray, fill=gray, fill opacity=0.2] (7,0) grid  (13,6) rectangle (7,0);
	\draw [very thin, draw=gray, fill=gray, fill opacity=0.2] (0,7) grid  (6,13) rectangle (0,7);
	\draw [very thin, draw=gray, fill=gray, fill opacity=0.2] (7,7) grid  (13,13) rectangle (7,7);
\end{scope}

\begin{scope}[canvas is xz plane at y=-10]
	\draw [very thin, draw=gray, fill=gray, fill opacity=0.2] (3,3) grid  (10,10) rectangle (3,3);
	\draw[fill=gray] (5,5) rectangle (6,6);
\end{scope} 
  \end{tikzpicture}

%% file: figures/nn_noiseless3D.tex
\begin{tikzpicture}
	\begin{axis}[
        height = 6.5cm,
		width = 6.5cm,
		xlabel=$p$,
        ylabel=$\overline{P}$,
		grid=major,
		xticklabel style={
		        /pgf/number format/fixed,
		        /pgf/number format/precision=5
		},
		scaled x ticks=false,
		yticklabel style={
		        /pgf/number format/fixed,
		        /pgf/number format/precision=5
		},
		scaled y ticks=false,
        xmin=0.10,
        xmax=0.26,
        legend cell align=left,
		legend pos=north west
		]

		\addplot[
			color=solarized-red,
			mark=diamond*,
			mark size=1,
			error bars/.cd,
			y dir=both,
			y explicit
		]
		table [x index=0, y index=1]
		{
0.0 0.0 0.0
0.025 0.0 0.0
0.05 0.0 0.0
0.075 0.0 0.0
0.1 0.000757575757576 0.0275136663596
0.1125 0.0028 0.0528408932551
0.125 0.01125 0.105467708328
0.1375 0.0356 0.185290690538
0.15 0.091 0.287609109731
0.1625 0.1835 0.38707589695
0.175 0.3168 0.465228717944
0.1875 0.4787 0.499546103978
0.2 0.6183 0.485803571415
0.2125 0.7308 0.443544090255
0.225 0.81555 0.387850741265
0.2375 0.8615 0.345424014799
0.25 0.89385 0.308029507515
		};

		\addplot[
			color=solarized-yellow,
			mark=square*,
			mark size=1,
			error bars/.cd,
			y dir=both,
			y explicit
		]
		table [x index=0, y index=1]
		{
0.0 0.0 0.0
0.025 0.0 0.0
0.05 0.0 0.0
0.075 0.0 0.0
0.1 0.0 0.0
0.1125 0.0006 0.024487547856
0.125 0.00225 0.0473807714163
0.1375 0.0102 0.10047865445
0.15 0.04905 0.215972446159
0.1625 0.1366 0.343424576872
0.175 0.30155 0.458930928899
0.1875 0.5064 0.499959038322
0.2 0.69 0.462493243194
0.2125 0.819183673469 0.384865928072
0.225 0.88615 0.317628993481
0.2375 0.92112244898 0.269547552327
0.25 0.9334 0.24932797677

		};

		\addplot[
			color=solarized-cyan,
			mark=triangle*,
			mark size=1,
			error bars/.cd,
			y dir=both,
			y explicit
		]
		table [x index=0, y index=1]
		{
0.0 0.0 0.0
0.025 0.0 0.0
0.05 0.0 0.0
0.075 0.0 0.0
0.1 0.0 0.0
0.1125 0.0003 0.0173179098046
0.125 0.0007 0.02644825136
0.1375 0.0039 0.0623280835579
0.15 0.0263636363636 0.160214216102
0.1625 0.1069 0.308986067647
0.175 0.285555555556 0.451678624961
0.1875 0.5504 0.497453354597
0.2 0.766464646465 0.423079888655
0.2125 0.8963 0.304870972708
0.225 0.94245 0.232890526858
0.2375 0.96 0.195959179423
0.25 0.9669 0.178897708202
		};

		\addplot[
			color=solarized-blue,
			mark=*,
			mark size=1,
			error bars/.cd,
			y dir=both,
			y explicit
		]
		table [x index=0, y index=1]
		{
            0.0 0.0 0.0
            0.025 0.0 0.0
            0.05 0.0 0.0
            0.075 0.0 0.0
0.1 0.0 0.0
0.1125 0.00030612244898 0.017493677087
0.125 0.0003 0.0173179098046
0.1375 0.00316326530612 0.0561538872984
0.15 0.00724489795918 0.0848080739832
0.1625 0.0779591836735 0.2681073467
0.175 0.2643 0.440959760069
0.1875 0.6328 0.482041657951
0.2 0.8364 0.3699122058
0.2125 0.9427 0.232414952187
0.225 0.9687 0.174127281033
0.2375 0.9765 0.151485147787
0.25 0.9855 0.119539742345
		};

	\legend{6,8,10,12}

%
%
%
	\end{axis}
\end{tikzpicture}

%% file: figures/failrate_global_3D.tex
\begin{tikzpicture}
    \large
    \begin{axis}[
        height = 6.5cm,
        width = 6.5cm,
        xlabel=$p$,
        ylabel=$\overline{P}$,
        grid=major,
        xmin=0.185,
        xmax=0.27,
        legend style={font=\small},
        legend pos=south east,
        legend cell align=left
        ]

        \addplot[
            color=solarized-cyan,
            mark=*,
            mark size=1,
        ]
        table [x index=0, y index=1]
        {
0.1 0.00337018064168 7.81279986525e-06
0.11 0.0048 1.38231110825e-05
0.12 0.0101265822785 1.81049007103e-05
0.13 0.0204 2.82728420927e-05
0.14 0.0327305605787 3.21755014303e-05
0.15 0.0502 4.36714826861e-05
0.16 0.0748643761302 4.75899420383e-05
0.17 0.1078 6.20255302275e-05
0.18 0.138628944505 7.52032448219e-05
0.19 0.192 9.84682690007e-05
0.2 0.223304070231 3.32370724083e-05
0.21 0.261916666667 3.66397937878e-05
0.22 0.2956 3.65049598822e-05
0.23 0.330833333333 3.92094304566e-05
0.24 0.3688 3.85983676339e-05
0.25 0.394833333333 4.07345720814e-05
0.26 0.42032 3.9488821515e-05
0.27 0.44175 4.13829454858e-05
0.28 0.46408 3.98966467644e-05
0.29 0.465333333333 4.15663978725e-05
0.3 0.47712 3.99580985013e-05
0.31 0.478 4.16263137931e-05
0.32 0.4832 3.99774144236e-05
0.33 0.48675 6.24780508334e-05
0.34 0.486235294118 5.88012349365e-05
0.35 0.495125 6.24970292263e-05
0.36 0.5 0.001
0.38 0.489361702128 0.00531794482774
        };

        \addplot[
            color=solarized-blue,
            mark=*,
            mark size=1,
        ]
        table [x index=0, y index=1]
        {
0.2 0.196629213483 4.96191731005e-05
0.21 0.2385 5.32707925955e-05
0.22 0.286875 5.65378377691e-05
0.23 0.335875 5.9036930229e-05
0.24 0.380625 6.06925666493e-05
0.25 0.41075 6.14962444865e-05
0.26 0.43675 6.19979129563e-05
0.27 0.4485 6.21675847508e-05
0.28 0.472125 6.24027974602e-05
0.29 0.487375 6.24800729952e-05
0.3 0.494 4.34751303221e-05
0.31 0.48775 6.24812393718e-05
0.32 0.49375 6.24951169968e-05
0.33 0.499375 6.24999511719e-05
0.34 0.501 6.24998749999e-05
0.35 0.48525 6.24727987682e-05
        };

\addplot[
color=solarized-magenta,
mark=*,
mark size=1,
            error bars/.cd, 
            x dir=both,
            x explicit,
            y dir=both,
            y explicit
]
table [x index=0, y index=1, y error index=2]
{
0.1 0.0 0.0
0.11 0.0 0.0
0.12 0.000443458980044 4.66824744046e-06
0.13 0.00075 6.84396403775e-06
0.14 0.00310421286031 1.23345720271e-05
0.15 0.00625 1.97023753327e-05
0.16 0.0155210643016 2.74086358765e-05
0.17 0.03725 4.73434509066e-05
0.18 0.0662971175166 5.51664618662e-05
0.19 0.10875 7.78313230245e-05
0.2 0.161310951239 2.9401880121e-05
0.21 0.22025 3.45345979398e-05
0.22 0.28288 3.60318885966e-05
0.23 0.332583333333 3.92615570142e-05
0.24 0.389826086957 4.24096244814e-05
0.25 0.426181818182 4.49564413639e-05
0.26 0.44947826087 4.31585741912e-05
0.27 0.462545454545 4.5326834865e-05
0.28 0.476 4.34281450297e-05
0.29 0.489545454545 4.54446082301e-05
0.3 0.494 4.34751303221e-05
0.31 0.501272727273 4.54543981966e-05
0.32 0.495043478261 4.34761245468e-05
0.33 0.489285714286 7.14121701286e-05
0.34 0.497466666667 6.66658109575e-05
0.35 0.495857142857 7.14261194914e-05
0.36 0.504 0.000999967999488
};

        \legend{4,5,6}
        
        
    \end{axis}

%
%
%
%
%
%


\end{tikzpicture}

%% file: figures/nn_noiseless4D.tex
\begin{tikzpicture}
	\begin{axis}[
        height = 6.5cm,
        width = 6.5cm,
		xlabel=$p$,
        ylabel=$\overline{P}$,
		grid=major,
		xticklabel style={
		        /pgf/number format/fixed,
		        /pgf/number format/precision=5
		},
		scaled x ticks=false,
		yticklabel style={
		        /pgf/number format/fixed,
		        /pgf/number format/precision=5
		},
		scaled y ticks=false,
        legend cell align=left,
		legend pos=north west
		]	
		

		\addplot[
			color=solarized-red,
			mark=diamond*,
            only marks,
			mark size=1,
			error bars/.cd,
			y dir=both,
			y explicit
		]
		table [x index=0, y index=1, y error index=2]
		{
			0.062 0.0325 0.00511889880736
			0.064 0.0583333333333 0.00676575664379
			0.066 0.0854545454545 0.00842895799209
			0.068 0.155454545455 0.0109248852606
			0.072 0.275 0.0141200212464
			0.074 0.372 0.0152845019546
			0.076 0.498 0.0158112618092
			0.078 0.598 0.0155047089621
		};

		\addplot[
			color=solarized-yellow,
            only marks,
			mark=square*,
			mark size=1,
			error bars/.cd,
			y dir=both,
			y explicit
		]
		table [x index=0, y index=1, y error index=2]
		{
			0.062 0.0158333333333 0.00360354442284
			0.064 0.0408333333333 0.00571299485254
			0.066 0.0636363636364 0.00736001339299
			0.068 0.134545454545 0.0102886952617
			0.07 0.197 0.0125774003673
			0.072 0.312 0.0146511433001
			0.074 0.424 0.0156276677723
			0.076 0.573 0.0156419627924
			0.078 0.714 0.014289996501
		};

		\addplot[
			color=solarized-cyan,
            only marks,
			mark=triangle*,
			mark size=1,
			error bars/.cd,
			y dir=both,
			y explicit
		]
		table [x index=0, y index=1, y error index=2]
		{
			0.062 0.0116666666667 0.00309980584816
			0.064 0.025 0.00450693909433
			0.066 0.0418181818182 0.00603545746459
			0.068 0.105454545455 0.00926056567571
			0.07 0.182 0.0122014753206
			0.072 0.313 0.0146639353517
			0.074 0.478 0.015796075462
			0.076 0.625 0.0153093108924
			0.078 0.798 0.0126962986732
		};

		\addplot[
			color=solarized-blue,
            only marks,
			mark=*,
			mark size=1,
			error bars/.cd,
			y dir=both,
			y explicit
		]
		table [x index=0, y index=1, y error index=2]
		{
			0.062 0.00583333333333 0.0021983527082
			0.064 0.0208333333333 0.00412303544689
			0.066 0.04 0.00590839156701
			0.068 0.0618181818182 0.00726114780652
			0.07 0.177 0.0120694241785
			0.072 0.331 0.0148808265899
			0.074 0.539 0.0157632166768
			0.076 0.714 0.014289996501
			0.078 0.859 0.0110054077616
		};

	\legend{6,7,8,9}

    \addplot [
    domain=0.0675:0.0745, 
    samples=100, 
    color=solarized-red,
    ]
    {0.263 + 2.784*(x-0.07126)*6^1.52 + 8.399 * ((x-0.07126)*6^1.52)^2};

    \addplot [
    domain=0.0675:0.0745, 
    samples=100, 
    color=solarized-yellow,
    ]
    {0.263 + 2.784*(x-0.07126)*7^1.52 + 8.399 * ((x-0.07126)*7^1.52)^2};
    
    \addplot [
    domain=0.0675:0.0745, 
    samples=100, 
    color=solarized-cyan,
    ]
    {0.263 + 2.784*(x-0.07126)*8^1.52 + 8.399 * ((x-0.07126)*8^1.52)^2};
    
    \addplot [
    domain=0.0675:0.0745, 
    samples=100, 
    color=solarized-blue,
    ]
    {0.263 + 2.784*(x-0.07126)*9^1.52 + 8.399 * ((x-0.07126)*9^1.52)^2};
	\end{axis}
\end{tikzpicture}

%% file: figures/failrate_global_4D.tex
\begin{tikzpicture}
    \large
    \begin{axis}[
        height = 6.5cm,
        width = 6.5cm,
        xlabel=$p$,
        ylabel=$\overline{P}$,
        xtick={0.05, 0.1, 0.15, 0.2},
        xticklabels={0.05, 0.1, 0.15, 0.2},
        grid=major,
        xmin=0.05,
        xmax=0.15,
        scaled x ticks=false,
        legend pos=south east,
        legend cell align=left
        ]


        \addplot[
            color=solarized-yellow,
            mark=*,
            mark size=1,
            error bars/.cd, 
            y dir=both,
            y explicit
        ]
        table [x index=0, y index=1]
        {
0.08 0.101166666667 5.02582144259e-05
0.09 0.186333333333 6.48959043011e-05
0.1 0.255833333333 3.63607158751e-05
0.11 0.3395 3.94616330898e-05
0.12 0.398666666667 4.08019909948e-05
0.13 0.439 4.13554208458e-05
0.14 0.46975 4.16324807676e-05
0.15 0.498 0.000499995999984
        };

        \addplot[
            color=solarized-cyan,
            mark=*,
            mark size=1,
        ]
        table [x index=0, y index=1]
        {
0.06 0.002 8.93532316148e-05
0.08 0.0390769230769 2.98120015385e-05
0.09 0.103833333333 5.08407022588e-05
0.1 0.21368 3.27922778812e-05
0.11 0.338083333333 3.9421422639e-05
0.12 0.42462230794 3.97207930396e-05
0.13 0.466916666667 4.15753577067e-05
0.14 0.496479570076 4.26356021962e-05
0.15 0.50 0.000499099188539
        };

        \addplot[
            color=solarized-blue,
            mark=*,
            mark size=1,
        ]
        table [x index=0, y index=1]
        {
0.08 0.00866666666667 6.17937629199e-05
0.09 0.058 0.000155828966071
0.1 0.182834777747 0.000108060109067
0.11 0.354570637119 0.00132516068139
0.12 0.45462230794 3.97207930396e-05
        };

        \legend{3,4,5}
        
        
    \end{axis}

%
%
%
%
%
%


\end{tikzpicture}

%% file: figures/nn_noisy4D_fit.tex
\begin{tikzpicture}
	\begin{axis}[
        height = 6.5cm,
        width = 6.5cm,
		xlabel=$p$,
        ylabel=$T$,
		xtick={0.027,0.03,0.032},
		xticklabels={0.027,0.03,0.032},
		grid=major,
		xticklabel style={
		        /pgf/number format/fixed,
		        /pgf/number format/precision=5
		},
		scaled x ticks=false,
		yticklabel style={
		        /pgf/number format/fixed,
		        /pgf/number format/precision=5
		},
		scaled y ticks=false,
		xmode = log,
        ymode = log,
        legend cell align=left,
		legend pos=south west
		]

		\addplot[
			color=solarized-red,
			mark=diamond*,
            only marks,
			mark size=1,
			error bars/.cd,
			y dir=both,
			y explicit
		]
		table [x index=0, y index=1, y error index=2]
		{
			0.0273 179.611398964 15.3256136743
            0.0277 150.222222222 12.6672671956
            0.029 82.0526315789 19.0787366431
            0.03 61.1578947368 10.727241265
            0.031 38.8421052632 9.03149945005
            0.032 20.8947368421 4.85840824075
		};

		\addplot[
			color=solarized-yellow,
            only marks,
			mark=square*,
			mark size=1,
			error bars/.cd,
			y dir=both,
			y explicit
		]
		table [x index=0, y index=1, y error index=2]
		{
			0.0273 335.459677419 35.0503023811
            0.0277 268.810218978 26.8424782879
            0.029 91.2631578947 21.2203523664
            0.03 33.9718309859 4.55359963295
            0.031 29.2407407407 4.41585557879
            0.032 14.4210526316 3.3531583324
		};

		\addplot[
			color=solarized-cyan,
            only marks,
			mark=triangle*,
			mark size=1,
			error bars/.cd,
			y dir=both,
			y explicit
		]
		table [x index=0, y index=1, y error index=2]
		{
			0.0273 528.830188679 80.5090070197
            0.0277 265.617977528 32.2148543133
            0.029 67.1052631579 15.6032002694
            0.03 33.8474576271 4.91926556469
            0.031 18.5142857143 3.36118373787
            0.032 12.7894736842 2.97378640429
		};

		\addplot[
			color=solarized-blue,
            only marks,
			mark=*,
			mark size=1,
			error bars/.cd,
			y dir=both,
			y explicit
		]
		table [x index=0, y index=1, y error index=2]
		{
			0.0273 596.6 124.360652614
            0.0277 570.482758621 111.945180044
            0.029 84.1071428571 16.7430101047
            0.03 30.8928571429 6.14976804269
            0.031 16.6315789474 3.86714610598
            0.032 10.5263157895 2.44756082657
		};

	\legend{5,6,7,8}

    \addplot [
    dashed,
    domain=0.0273:0.032, 
    samples=100, 
    color=solarized-red,
    ]
    {(x/exp(-3.1976327807303027))^(-12.943050123486122)};

    \addplot [
    dashed,
    domain=0.0273:0.032,
    samples=100, 
    color=solarized-yellow,
    ]
    {(x/exp(-3.3139157219305853))^(-20.12870109435307)};
    
    \addplot [
    dashed,
    domain=0.0273:0.032, 
    samples=100, 
    color=solarized-cyan,
    ]
    {(x/exp(-3.3459521947955815))^(-23.31603152410292)};
    
    \addplot [
    dashed,
    domain=0.0273:0.032, 
    samples=100, 
    color=solarized-blue,
    ]
    {(x/exp(-3.369082824999092))^(-27.542006948811572)};
	\end{axis}
\end{tikzpicture}

%% file: discussion.tex
\label{sec:discussion}

We have shown how convolutional neural networks can be utilized to decode higher dimensional topological quantum codes, leading to a scalable architecture. 
The neural network is trained once and can then be scaled up to arbitrary system sizes.
However, our approach is only one way of utilizing neural networks to decode high-dimensional quantum codes.
Given the versatility of neural networks, it is clear that there will be other approaches to the problem, which are worth exploring.
As we mentioned in \autoref{sec:ML_for_decoding}, ideally we need to perform reinforcement learning (i.e. optimize the parameters of the neural networks with the objective being lowest logical error rate or longest memory time).

There is another important reason that we choose neural networks instead of other machine learning models for decoding.
As discussed in~\cite{bengio2007scaling}, good learning algorithms should be efficient in terms of human involvement.
For example, it is highly undesirable if small changes in the quantum error correcting code and the experimental hardware require a large amount of human effort to rewrite the learning algorithms.
These environmental changes are certainly going to happen very frequently before a large-scale quantum computer is built.
On the other hand, if a certain class of learning algorithms are fast to implement, then with the same amount of man-hour we can test more different algorithms for the problem.
Therefore, less human involvement often translates to better performance.
At the moment of writing, neural networks have some of the most flexible and automated software packages among machine learning models.

We would also like to highlight that using neural networks for decoding may have advantages from a practical perspective.
Many specialized neural network chips have been manufactured~\cite{Mnnhardware,Mintegratedcircuit,jouppi2017datacenter} which promise much lower power consumption.
If the decoding has to take place inside a fridge such dedicated hardware could help to keep down thermal noise.
For example, in \cite{Mintegratedcircuit} the authors report on an integrated circuit implementing a network of 1 million neurons on a $240\mu \mbox{m} \times 390 \mu \mbox{m}$ CMOS chip.
The chip draws $20 \mbox{mW}/\mbox{cm}^2$ as compared to $50\mbox{W}/\mbox{cm}^2$-$100\mbox{W}/\mbox{cm}^2$ for a modern CPU or $30\mbox{W}/\mbox{cm}^2$ for an FPGA \cite{Thardwarecomparison}, thereby reducing potential thermal noise by orders of magnitude.

A natural question is whether we can build a similar convolutional neural network decoder for 2D toric code.
As the architecture we proposed heavily based on the fact that 4D toric code can be decoded in a local single-shot manner, it cannot be directly applied to 2D toric code.
However, we foresee that a decent convolutional neural network decoder for 2D toric code exists, and it will likely share many similarities to the renormalization group decoder in~\cite{duclos2013fault}.


\subsection*{Acknowledgments}
We would like to thank Christophe Vuillot and Kasper Duivenvoorden for interesting discussions and Barbara Terhal for feedback on our manuscript.

%% file: training_neural_network.tex
\label{appendix:training_neural_network}
We have mentioned in the main text that neural networks are a powerful ansatz to model functions $F : \{0,1\}^{m} \rightarrow [0,1]^{n}$.
The question is how to choose the individual weights and biases of the neurons to make the network compute $F$, or at least give a good approximation.
This task can be formulated in terms of an optimization problem where pairs of input and desired output $(x,F(x))$ are used to find the right weights and biases.
In our setup we assume that the inputs of $F$ are weighed by some probability distribution $P:\{0,1\}^{m} \rightarrow [0,1]$.
The distribution $P$ prioritizes certain inputs over others and effectively reduces the dimensionality of the input space.
In principle we would want to optimize over all possible pairs of inputs and outputs of $F$ (while taking $P$ into account).
However, this is generally not practicable so that we restrict ourselves to optimize over some subset $D \subset \{ (x,F(x)) \mid x \in \{0,1\}^m\}$. 
The set $D$ is sampled according to this distribution $P$.
The optimization of the network is called {\em training} and the sample $D$ is called the {\em training data} or {\em training set}.

We will now describe the training of neural networks based on gradient descent.
We denote the weight vector of the $i$th neuron in layer $l$ by $w_i^l$ and the $j$th entry of this vector by $w_{i,j}^l$.
Similarly, the bias of the $i$th neuron in the $l$th layer is labeled $b_i^l$.
These are the parameters that we need to optimize.
An essential ingredient for the optimization is a measure of how good a neural network performs on the training data $D$.
This measure is called the {\em cost function} $C_D(w_{i,j}^l,b_i^l)$ which usually maps the values of the weights $w_{i,j}^l$ and biases $b_i$ of the neural network into~$[0,\infty]$.
If the value of the cost function is small then this is an indicator that the network performs well.
For reasons that will become apparent in the following discussion, we demand $C_D$ to be differentiable.
An obvious choice for the cost function is the average squared $L^2$ norm $\Vert \cdot \Vert^2$ of the difference of the networks output $F_{N}(x,w_{i,j}^l,b_i^l)$, which depends on the choice of the weights $w_{i,j}^l$ and biases $b_i^l$, and the desired value $F(x)$ over all elements of the training set~$D$:

\begin{align}\label{eqn:L2cost_function}
C_D(w_{i,j}^l,b_i^l) =  \frac{1}{2 |D|} \sum_{(x,F(x)) \in D} \Vert F^N(x,w_{i,j}^l,b_i^l) - F(x) \Vert^2
\end{align}

To optimize the weights and biases we perform an iterative procedure called {\em gradient descent}.
A good introduction to gradient descent and its variants can be found in~\cite{goh2017why}.
Generally, gradient descent is a tool to find a local minimum of a differentiable function $f:\real^n \rightarrow \real$ which is close to some initial point $x_0\in \real^n$.
In the first step we evaluate the negative gradient $-\nabla f$ at $x_0$.
By following the negative gradient for a small enough distance we will obtain a point $x_1 := x_0 - \eta_0 \nabla f (x_0)$ such that $f(x_1)\leq f(x_0)$.
Iterating this process gives a sequence $x_0,x_1,x_2,x_3,\dotsc$, where
\begin{align}\label{eqn:graddescent}
x_{i+1} := x_i - \eta_i \nabla f(x_i).
\end{align}
If the parameters $\eta_i$ where chosen small enough we have that $f(x_{i+1})\leq f(x_i)$ so that the sequence $x_i$ will converge towards the location of a local minimum.
Clearly, we do not want to choose the $\eta_i$ too small or otherwise the rate of convergence of the $x_i$ will be slow.
However, if we are choosing $\eta_i$ too large it will make us overshoot the location of the local minimum.
There are situations in which $f$ satisfies conditions, i.e. if $f$ is convex and smooth, in which there exists an explicit choice of $\eta_i$ for which the convergence can be guaranteed.
In the context of training neural networks the parameters $\eta_i$ are collectively referred to as the {\em learning rate}. 
At the time of writing there is no developed theory on how to choose the learning rates optimally and we have to consider heuristics.
An overview of several heuristics can be found in \cite{orr2003neural}.

Let us now apply gradient descent to optimize neural networks.
The setup is the following:
We have some set of training data $D$, a neural network with some initial choice of weights and biases and a cost function $C_D$.
The task is to find weights and biases which (locally) minimize the cost function $C_D$.
This confronts us with the problem of how to compute the gradient $\nabla C_D$.
This is solved by the second major ingredient of the training of neural networks: The {\em backpropagation algorithm}.
The backpropagation algorithm consists of two steps:
In the first step we compute the cost function $C_D$ of the neural network (Eqn.~\ref{eqn:L2cost_function}) as well as record all values of the neurals in the network.
To evaluate the output of the network $F^N$ we evaluating the input $x$ on the first hidden layer and then feed the output of the first hidden layer into the second hidden layer and so forth until we obtain the output of the network at the last layer.
In the second step of the backpropagation algorithm we compute the derivative of the cost function with respect to all weights and biases.
The derivatives can be computed in linear time in the number of neurons.
Obtaining the derivatives is a matter of applying the chain rule several times.
To simplify notation we introduce the variable $s_i^l = \sum_{k\in pred(i,l)} w_{i,k}^l f_k^{l-1}+b_i^l$, where $pred(i,l)$ and $f_i^{l}$ are the predecessors and the value of the $i$th neuron in the $l$th layer respectively.
Remember that the value of a neuron is $f_i^{l} = \sigma(s_i^l)$.
The derivatives of the cost function $C_D$ with respect to the weight $w_{i,j}^l$ can be expanded as
\begin{align}\label{eqn:backpropagation}
\begin{split}
    \frac{\partial C_D}{\partial w_{i,j}^l} &= 
    \frac{\partial C_D}{\partial s_i^l}
    \frac{\partial s_{i}^l}{\partial w_{i,j}^l}
\end{split}
\end{align}
The second factor of \autoref{eqn:backpropagation} is simply
\begin{align}\label{eqn:backpropsum}
\frac{\partial s_{i}^l}{\partial w_{i,j}^l} = f^{l-1}_i .
\end{align}
The form of the first term of \autoref{eqn:backpropagation} depends on whether $l$ is a hidden layer or the output layer.
For $l=L$ we expand over the values of the neurons in the output layer $F^N_k = f_k^L$
\begin{align}\label{eqn:backprop1}
\frac{\partial C_D}{\partial s_i^L} = \sum_{k=1}^{n} \frac{\partial C_D}{\partial f_k^L} \frac{\partial f_k^L}{\partial s_i^L} = \frac{\partial C_D}{\partial f_i^L} \sigma'(s_i^L)
\end{align}
For all hidden layers $l<L$ we expand over the sums~$s_k^{l+1}$ of neurons which are in the next layer and connected to the $i$th neuron in layer $l$
\begin{align}\label{eqn:backprop2}
\begin{split}
\frac{\partial C_D}{\partial s_i^l} &= \sum_{k\in succ(i,l)} \frac{\partial C_D}{\partial s_k^{l+1}} \frac{\partial s_k^{l+1}}{\partial s_i^l}\\
&= \sum_{k\in succ(i,l)} \frac{\partial C_D}{\partial s_k^{l+1}}\; w_{i,k}^{l+1} \sigma'(s_i^l)
\end{split}
\end{align}
where $succ(i,l)$ indicates the set of all neurons in the next layer connected to the $i$th neuron in layer $l$.
The derivatives with respect to the biases $b_i^l$ proceeds completely analogously, the only difference being that \autoref{eqn:backpropsum} evaluates to 1.

Note that in order to compute \autoref{eqn:backprop2} for some layer $l$ we need to have the result of layer $l+1$.
Hence we first evaluate \autoref{eqn:backprop1} for the output layer and then go backwards through the layers of the network (hence the name backpropagation).
Finally, having obtained all derivatives with respect to the weights and biases allows us to perform a single step in the gradient descent (see \autoref{eqn:graddescent}).

In the discussion above we made two simplifications which we are going to address now: 
The first simplification was the choice of the cost function. 
The $L^2$ norm is very intuitive but it leads to a very slow convergence of the training.
The reason for this can be seen in \autoref{eqn:backprop1} and \autoref{eqn:backprop2}.
The derivatives are proportional to the derivative of the sigmoid function $\sigma'(z)$ which is close to 0 when $|z|$ is sufficiently large.
This is avoided by choosing the {\em cross-entropy} as cost-function:
\begin{widetext}
\begin{align}\label{eqn:crossentropy}
    C_D = -\frac{1}{|D|} \sum_{x\in D}\left(  \sum_{k=1}^{n} \left[ F_k(x) \log\left(F_k^N(x)\right) + (1-F_k(x)) \log\left(1 - F_k^N(x)\right) \right] \right)
\end{align}
\end{widetext}
The cross-entropy has a less intuitive form.
However, since $F(x)\in [0,1]$ and $0<F^N(x)<1$ one can see that the cross-entropy is (a) positive and (b) small when the output of the network is close to the desired output.
The cross-entropy has the advantage that in Eqn.~\ref{eqn:backpropagation} the derivatives of the sigmoid function cancel, see \cite{Gmachinelearning,Mmachinelearning,Nmachinelearning} for a derivation.

The second simplification was taking the average over the whole training set $D$.
In practice, the training set is usually subdivided into several subsets called {\em batches} so that a batch of data can be loaded into memory.
This is known as \textit{stochastic gradient descent}.
Furthermore, part of the available training data is kept aside and not used for the training of the network.
This data set is the called the {\em validation set} $V$ and it is only used to evaluate the cost function after every step of the training.
The reason to keep the validation set separate is to check whether the neural network performs well on data outside of the training data.
In other words, this is the first measure against overfitting
To summarize the training procedure:
\begin{enumerate}
	\item Initialize the weights and biases.
	\item Pick a batch $B\subset D$ and learning rate $\eta$.
	\item Perform a single step of the gradient descent, using backpropagation to compute the gradient.
	\item Compute the cost function $C_V$ on the validation set.
\end{enumerate}
As long as $C_V$ keeps descending we repeat steps 2 - 4.
The initial values in step 1 are usually chosen to be random.

%% file: parallel_line_decoder.tex
\label{sec_parallel_line_decoder}

In this section, we will describe in detail the parallel line decoder.
As we mentioned in \autoref{subsection:architecture_and_training}, its purpose to decode the syndromes that consist only a few parallel lines, and we designed it with easiness to code in mind. The steps are the following:
\begin{enumerate}
	\item Make a list of current violated parity checks. Order the violated edges in the list by their direction.
	\item For each edge $e$ in the list: Assume $e$ has coordinate $(x_0,x_1,x_2,x_3,d)$, we look for the closest edge $e' = (y_0,y_1,y_2,y_3,d)$ that still has a violated parity check, such that $x_d=y_d$. We then change one of the $x_i$ in $e$ so that $e$ and $e'$ get closer by flipping the corresponding qubit. Update the syndrome.
	\item Repeat Step $1,2$ until no parity check is violated or certain time limit is exceeded. 
\end{enumerate}

\begin{figure}[htb]
	\centering
	\includegraphics[width=0.6\linewidth]{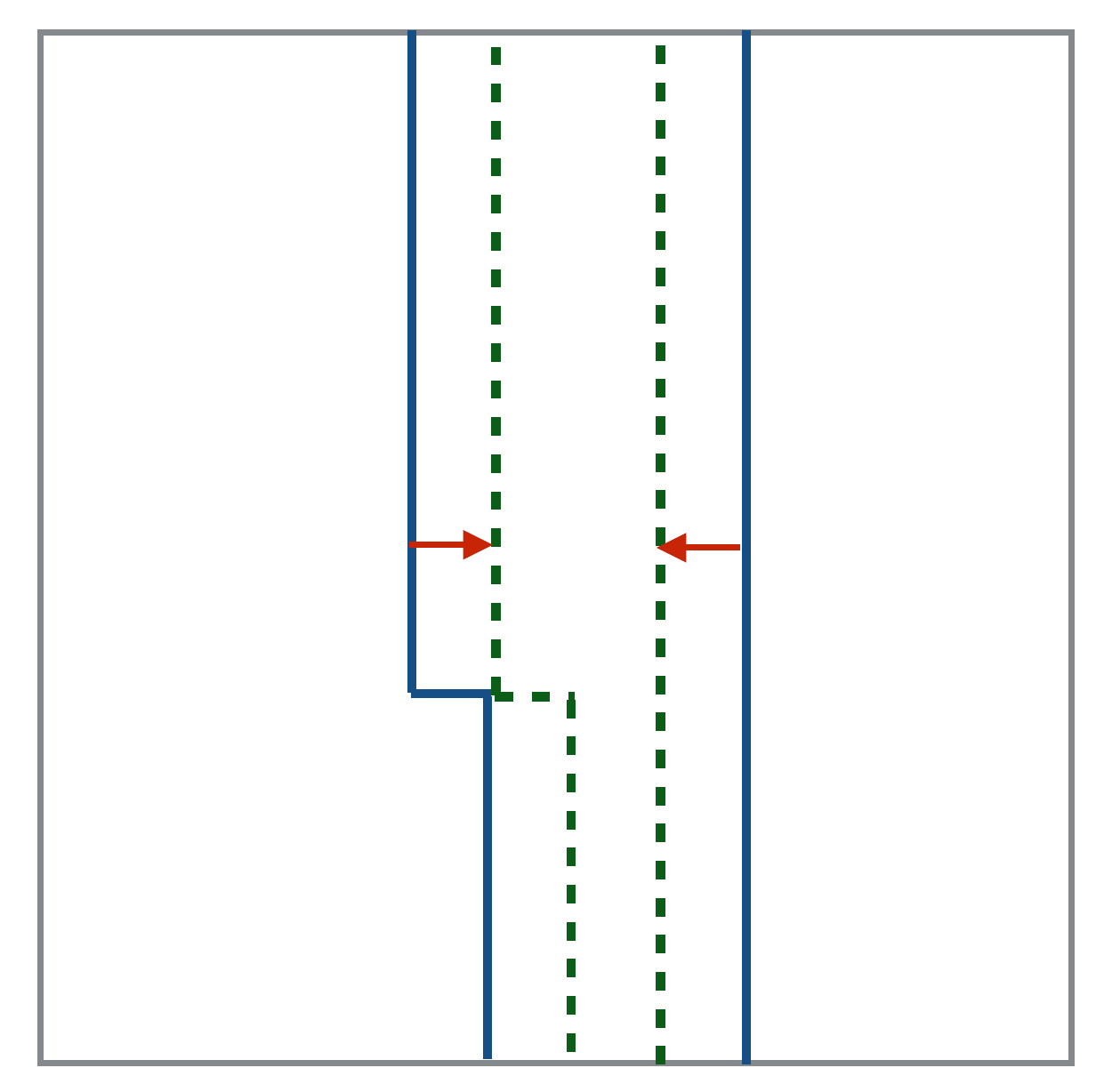}
	\caption{The effect of Step 2 illustrated in a 2D cross section. The solid blue lines are the initial violated parity checks, and the dotted green lines are the violated parity checks after a single execution of the Step 2. In the next loop, no violated parity check will be left.}
	\label{fig:parallel_line_decoder}
\end{figure}

%% file: alphago.tex
\label{appendix:alphago}

 In this section we will give a brief summary about the architecture of AlphaGo~\cite{silver2016mastering}, which ends up being a strong AI at playing board game Go. It highlights the importance of combining different type of machine learning in solving complicated tasks.
\begin{enumerate}
	\item Supervised learning is used to train a model to approximate human strategy, where the optimization is done to predict human move with most accuracy. While solely mimicking human moves can already achieve non-trivial performance~\cite{clark2015training}, it is almost surely not the best strategy. Therefore the following steps are needed.
	\item Reinforcement learning is then used to further improve the model, with the goal of achieving best win rate against previous trained models. Roughly speaking, it is done by gradually changing parameters of the neural network towards the direction of winning more games, starting from the parameters obtained from the supervised learning above.
	\item Monte-Carlo tree search is hand-picked as a subroutine, which reflects the 2-player turn by turn nature of the game Go. The details of Monte-Carlo tree search is not of concern to this paper. Here the point being a large and important fraction of the AlphaGo AI is pre-determined.
\end{enumerate}
While it is possible with only the reinforcement learning in the step 2, the trained AI can still achieve a similar performance as the AlphaGo, it is a much riskier approach. This is because the reinforcement learning can get stuck at some local minimum. The goal of step 1 is to set the start point of optimization to be approximately the human strategy, so that we can hope the local minimum obtained by reinforcement learning is at least better.

In a later paper~\cite{silver2017mastering}, the authors proposed a new training procedure which does not require the dataset of human moves.
With the nature of 2-player game in mind, they use the policy neural network which decides the next move to play out the following few turns of both players.
By doing that, they are able to recognize the better next move, and use that as a training target for the policy neural network.
Overall, this approach avoids the difficulty of training the policy neural network solely based on the win or loss of a match which typically consists of hundreds of moves.
Similarly, in our paper we try to avoid the same difficulty by explicitly training the neural network to recognize qubits affected by errors from the syndromes, instead of training the neural network to achieve longer memory times.
We envision the possibility of applying reinforcement learning after the supervised learning phase for our decoder, so that it can even find a slightly better strategy or adapt to not drastically different noise models.

%% file: lookup_table.tex
\label{appendix:nearest_neighbor}

The goal of this section is to demonstrate a baseline machine learning algorithm can fail at a toy problem similar to decoding topological codes when the lattice gets large.
This should serve as an alert when attempt to use neural networks for decoding, especially due to the lack of understanding of the generalization behavior of neural networks (see~\cite{zhang2016understanding}).
The simple algorithm we will be considering is the $1$-nearest neighbor algorithm.
It belongs to the family of $k$-nearest neighbor algorithms, which is used for classification and regression. The output of these algorithms only depend on the $k$ entries in the dataset that are closest to the input, e.g. average over the $k$ entries for regression.
It can be considered as a natural extension to the lookup table when the input space is too large and we cannot pre-compute all possible inputs (lookup table has been used for decoding the surface code in~\cite{tomita2014low})
The concrete procedure of using 1-nearest neighbor algorithm for our toy problem will be explained below.

Suppose we are given a square lattice with size $L^2$. Among all the plaquettes, $\lfloor p L^2 \rfloor$ are flipped, where $p<0.1$ is some fixed small number.
Our goal is to flip them back, but it is forbidden to look at the lattice directly, and we can only compare it to entries of a database.
The database contains $N=\text{poly}(L)$ entries, and each one is generated by randomly flip $\lfloor p L^2 \rfloor$ plaquettes.
A natural way to use the database is to find the entry that has the most overlapping flipped plaquettes with our lattice, and then flip all the plaquettes in that entry.
This approach can be considered as a $1$-nearest neighbor algorithm.
Intuitively, this strategy will perform poorly if $L$ is large, as the entries of the database are too sparse compared to all possibilities.
In more detail, define the random variable $C$ to be the number of overlapping flipped plaquettes between two randomly generated lattice configuration described above.
For large $L$, $C$ can be well approximated by $\mathcal{N} (L^2 p^2, L^2 p \alpha)$. Or equivalently,
\begin{equation}
\frac{C-L^2p^2}{L\sqrt{p \alpha}} \approx \mathcal{N}(0,1)
\end{equation}
Since it is known that~\cite{david2004order} 
\begin{align}
\operatorname{E}\left(\max_{i\leq N} (X_i) \right) & \sim \sqrt{\log N} \\
\operatorname{Var}\left(\max_{i\leq N} (X_i) \right) & \sim 1/\log N
\end{align}
where $X_i \sim \mathcal{N}(0,1)$. Thus, in the database, the most similar entry will typically have around $L^2 p^2 + L \sqrt{p\log N} \alpha'$ overlapping plaquettes. This is much smaller compare to $L^2 p$. Therefore, if we flip plaquettes according the entry, the total number of flipped plaquettes will increase instead of decrease.

Instead of this toy problem, we can apply a similar 1-nearest neighbor algorithm to a dataset of (syndrome, position of flipped qubits).
Let us consider the regime where the error rate is low.
In this case, most flipped qubits are separated from each other, and the syndromes in the dataset are almost equivalent to a list of flipped qubits.
Therefore, we might extrapolate from the toy problem discussed above that this particular 1-nearest neighbor algorithm will fail to decode topological codes when the lattice becomes large enough.
Thus, if we want to have a machine learning decoder that is scalable, it should ideally distinguish from such an approach.

%% file: analysis_cnn.tex
\label{appendix:analysis_larger_network}

In the paper, the convolutional neural networks we use only have one convolution layer with kernel size~$3^4$ and others are $1\times 1 \times 1 \times 1$. Therefore each element in the output (before applying softmax) is determined by its surrounding $3^4$ region. This is mainly for the sake of fast training and evaluation.
It is  interesting to ask what will happen if we have a deeper neural network, especially considering the recent success of deep neural networks in image classification and processing.
More concretely, we wish to know if we increase the number of convolution layers (and as a result increase the region each qubit can ``look around'' to decide its error probability), will the machine learning procedure described in our paper perform better?
In the previous section we give an example that a nearest neighbor algorithm performs worse on a larger region. And in theory a deeper neural network will have a better ability to remember all the training data, thus has the capacity to be a nearest neighbor algorithm.
Therefore, it is important to try to understand some difference between how neural nets and nearest neighbor algorithm generalize to unknown inputs.
To this end, we tested a slightly larger neural network for our decoder.

\subsection{Behavior of a deeper neural network trained as decoder}
\label{appendix:behavior_larger_network}
\begin{figure}
	\centering
	\includegraphics[width=0.9\linewidth]{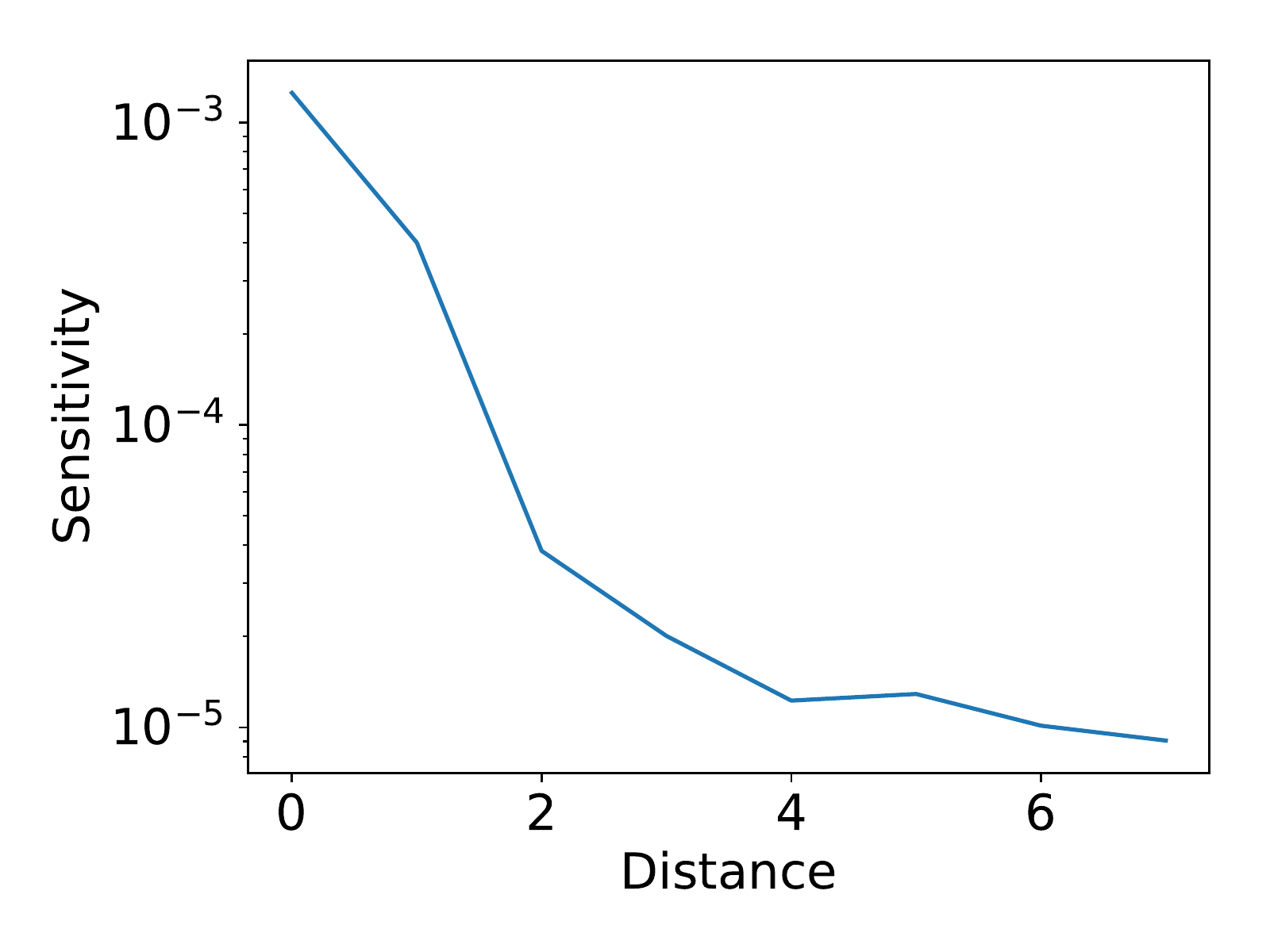}
	\caption{Sensitivity of the neural network output with respect to changing a single input bit. The neural network is trained to estimate the probability of an error happened on the center qubit. The X-axis is the distance between the single changed input bit and the center, induced by $L^1$-norm. For each distance, we pick a random input bit and evaluate the difference of output if we change that bit. This is repeated for 20 times and an average is computed.}
	\label{fig:sensitivity_vs_dist}
\end{figure}

In this subsection, We consider a neural network which has two convolution layer with kernel size $3^4$.
Thus, each element in the output of the neural network can be influenced by a $5^4$ region, or around 2500 input bits.
Generally speaking, if we have large neural nets on this many input bits, overfitting is very likely to happen.
For our particular setup of convolutional neural networks, it is much harder to estimate how many training data are needed to avoid overfitting.
Nevertheless, it is still interesting to look at the following two facts:
\begin{itemize}
	\item The trained larger network has a similar performance when decoding noiseless syndromes compared to the network we use in the main text, but is not tested as extensively.
	\item The larger network exhibits the decay of sensitivity, which means each element in the output will on average change more if an input bit closer to it is being changed. See \autoref{fig:sensitivity_vs_dist} for the plot.
\end{itemize}
Note that the second fact partially explained the first one: if the input bits outside the $3^4$ region do not have much influence on the corresponding output bit, then the smaller neural network used in the paper can in principle approximate the same input-output relation.
The decay of sensitivity is likely resulted from the following two reasons.
First, the decoding task, and thus the dataset we train our neural networks on have this structure.
In more detail, measurement outcome of a parity check far away from a qubit contains very little information about whether an error happened on the qubit.
Secondly, as we will discuss in the next subsection, a convolutional neural network has some tendency of prioritizing local information.
Intuitively, the alignment of the tendency of the convolutional neural network and the decoding task will lower the chance of overfitting.
Therefore, we might think convolutional neural network is not only a convenient machine learning model for this task, but also a naturally suitable one.

\subsection{Tendency of using local information}
As mentioned in the last subsection, here we will discuss one mechanism that make the convolutional neural network prioritize local information.
The mechanism mainly involves the local connectivity, the initialization and training of the networks.
We will explain the mechanism using a highly simplified setting.
However, we believe it still has some effect for more general cases.
We do not aim for rigorousness in this subsection.

\begin{figure}[htb]
	\centering
	\includegraphics[width=0.5\linewidth]{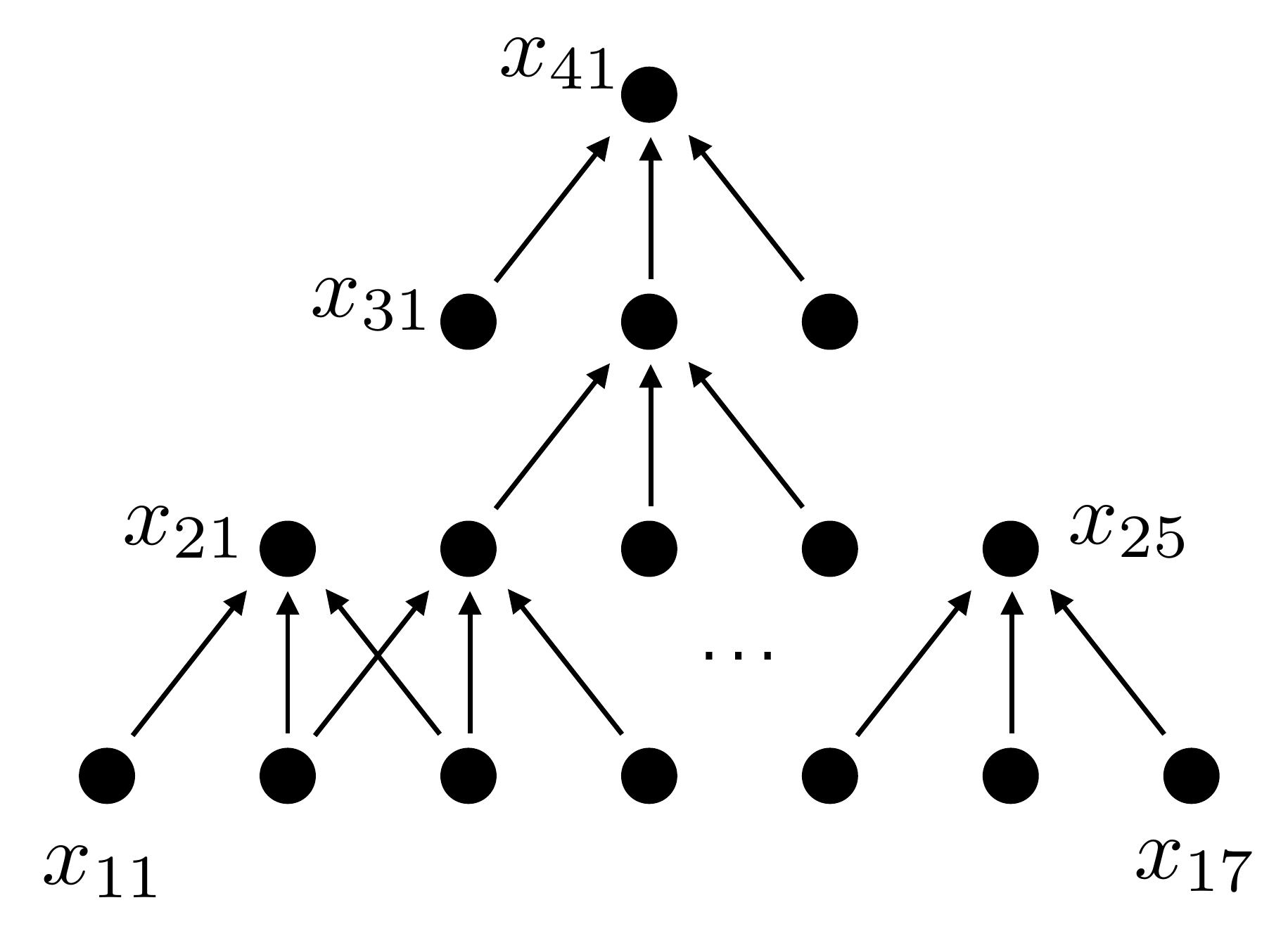}
	\caption{An illustration of the connectivity of a small 1D convolutional neural network. In this section, we set the number of channels to $1$. Thus, each $x_{ij}$ is a real number. The arrays represent dependence relation. For example, $x_{21}$ is a function of $x_{11},x_{12},x_{13}$, and in this section the function is simply a linear sum.}
	\label{fig:1d_conv}
\end{figure}

In \autoref{fig:1d_conv} we draw a $m=4$ layer 1D CNN with $n=7$ input bits and 1 output bit. The variables $x_{ij}$ are the values of the neurons (as well as inputs). 
We will first consider the following simplified situation:
\begin{enumerate}
	\item  There is no non-linear activation function applied in the middle of the network. The function $\tanh$ is only applied after the final layer $x_{m1}$ (In the figure this is $x_{41}$ ). There is also no bias term in the network. So the last layer $x_{m1}$ is a linear combination of the input bits $\{x_{1i}\}$.
	For the ease of notation, the number of channels for each layer is set to $1$.
	\item  The neural network is not convolutional, but instead a normal NN with the same local connectivity as a CNN.
	\item  The input-output relation to be learned as $y=x_{11}=x_{1k}=\pm 1$, where $k=\lfloor n/2 \rfloor$, $y$ is the desired output, and the probabilities of being $1$ and $-1$ are both $0.5$. Other $x_{1i}$ are set to $0$. The cost function is $c=(y-\tanh x_{m1})^2$.
\end{enumerate}

It is obvious that the neural net can approximate the above input-output relation well by being a linear combination $x_{m1}=a_1 x_{11}+a_k x_{1k}$ with the weights satisfy $a_1+a_k \gg 1$.
We will argue that with gradient descent as training method, in general we will have $a_k> a_1$. In other words, the output of the network $y'=\tanh (x_{m1})$ will depends more heavily based on $x_{1k}$ compared to $x_{11}$. Operationally, this can be checked by setting $x_{11}\neq x_{1k}$.
As we never train the neural net with data that have $x_{11}\neq x_{1k}$, this can be considered as checking how the neural net generalize to unseen inputs.

\begin{figure}
	\centering
	\includegraphics[width=\linewidth]{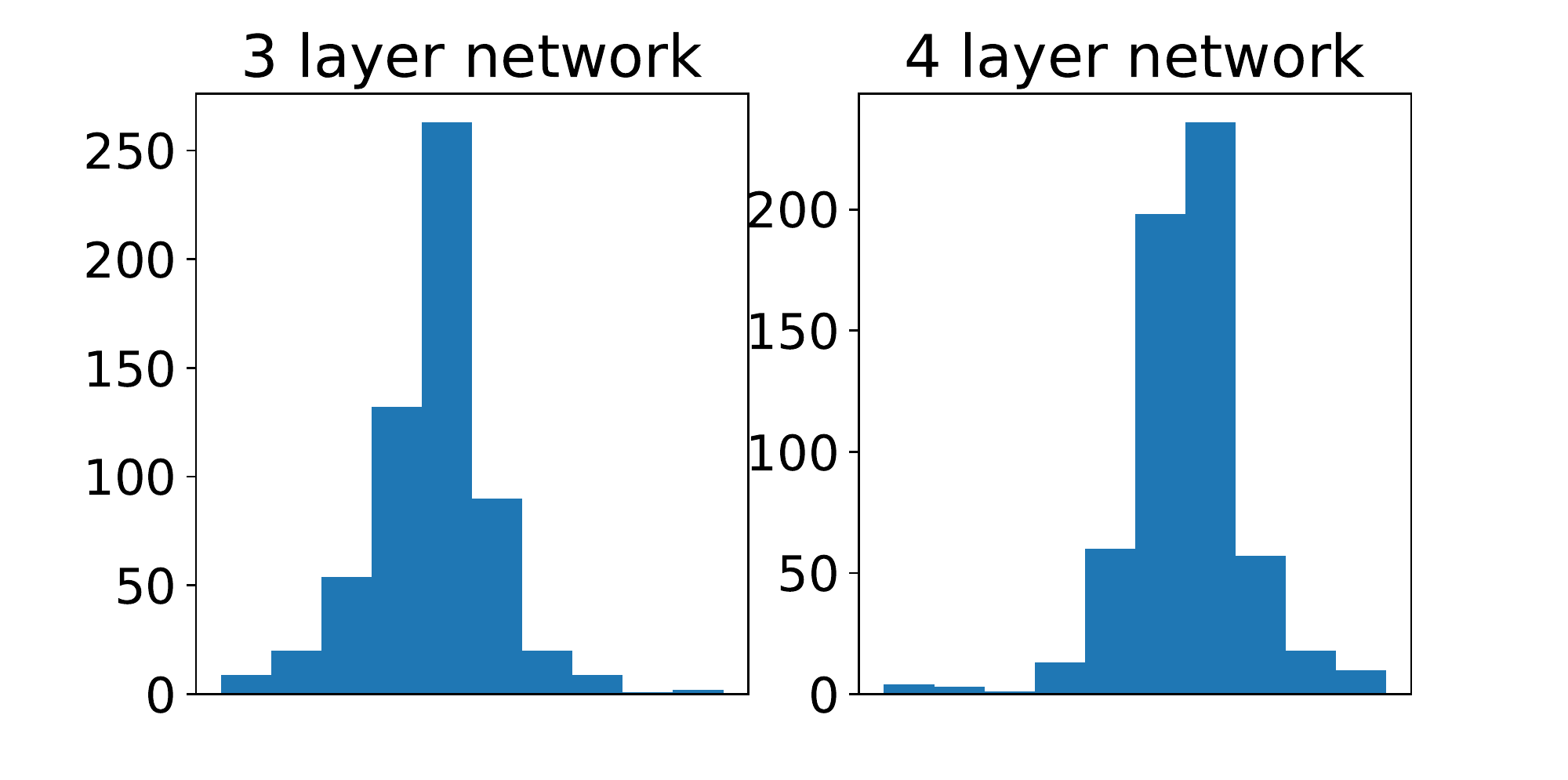}
	\caption{The distribution of the outputs from 3 and 4 layer neural network when the weights are initialized. 600 numerical trials are done in order to plot each figure. The networks have the connectivity shown in \autoref{fig:1d_conv}. The only non-zero element we set in the input is the center element $x_{1 \lfloor n/2 \rfloor}=1$. Thus, what we are showing here is the distribution of $a_{\lfloor n/2 \rfloor}^{(m1)}$, where $a_{k}^{(ij)}$ is defined in \autoref{eq_linear_network_coefficients} and $x_{m1}$ is the output of the network.
	We see that the distribution can be described as bell shapes.
    Thus, the variances already contain a large amount of information of the distribution.}
	\label{fig_pathsum}
\end{figure}

We will assume that the weights $\{w_{i}\}$ are initialized with the same normal distributions with mean $0$ and standard deviation $\sigma$. Since there is no non-linear activation function, we have
\begin{align}
x_{ij}=\sum_k a^{(ij)}_k x_{1k} 
\label{eq_linear_network_coefficients}
\end{align}
We will show that when the weights of the network are initialized, $\operatorname{Var}(a^{(ij)}_k)$ is proportional to the total number of paths from $x_{1k}$ to $x_{ij}$.
Let $p$ be any path from $x_{1k}$ to $x_{ij}$. We can think $p$ as the set of weights $w_i$ on the path. It is easy to show that
\begin{align}
	a^{(ij)}_k = \sum_p \prod_{w_i \in p} w_i \equiv \sum_p W_p
\end{align}
Note that for different path $p_1$, $p_2$, while $W_{p_1}$ is not independent of $W_{p_2}$, we have $\operatorname{Cov}(W_{p_1},W_{p_2})=0$.
Thus, 
\begin{align}
	\operatorname{Var}(a^{(ij)}_k )=  \sum_p \operatorname{Var}(W_p)
\end{align}	
More generally, we can formulate the above argument as the following observation:
\begin{observation}
	\label{obs_variance_prop_num_paths}
	Define $b_{(i_1j_1)}^{(i_2j_2)}=\sum_{p\in S} \prod_{w_k \in p} w_k$, where $S$ is the set of paths from $x_{i_1j_1}$ to $x_{i_2j_2}$.
	When the weights of the network are initialized, $\operatorname{Var}(b_{(i_1j_1)}^{(i_2j_2)})$ is proportional to the total number of paths in $S$.
\end{observation}
We want to use the variance to compare the magnitudes of $b_{(i_1j_1)}^{(i_2j_2)}$.
In order for this to work, one condition is that the distributions of $b_{(i_1j_1)}^{(i_2j_2)}$ should all have a ``regular shape'' (e.g. approximated by normal distribution).
In \autoref{fig_pathsum}, we plotted the distribution of $a_{\lfloor n/2 \rfloor}^{(m1)}$, which is defined in \autoref{eq_linear_network_coefficients}. Again $m$ is the layer of the network and $n$ is the size of input.
From this numerical experiment, it is likely that the probability distribution of $b_{(i_1j_1)}^{(i_2j_2)}$ will have a bell shape when $x_{i_1j_1}$ and $x_{i_2j_2}$ are reasonably far away.
Thus, the variances will provide us a very rough way to compare the magnitudes of $b_{(i_1j_1)}^{(i_2j_2)}$.
For example, this implies that in \autoref{fig:1d_conv}, $x_{32}$ likely has a larger dependence on the $x_{14}$ compared to $x_{31}$ on $x_{11}$.
Some other conditions will be needed if we want to make the above comparison rigorous.
However, we will skip further discussion on this topic and be contend with an incomplete analysis.

\begin{figure}
	\centering
	\includegraphics[width=0.75\linewidth]{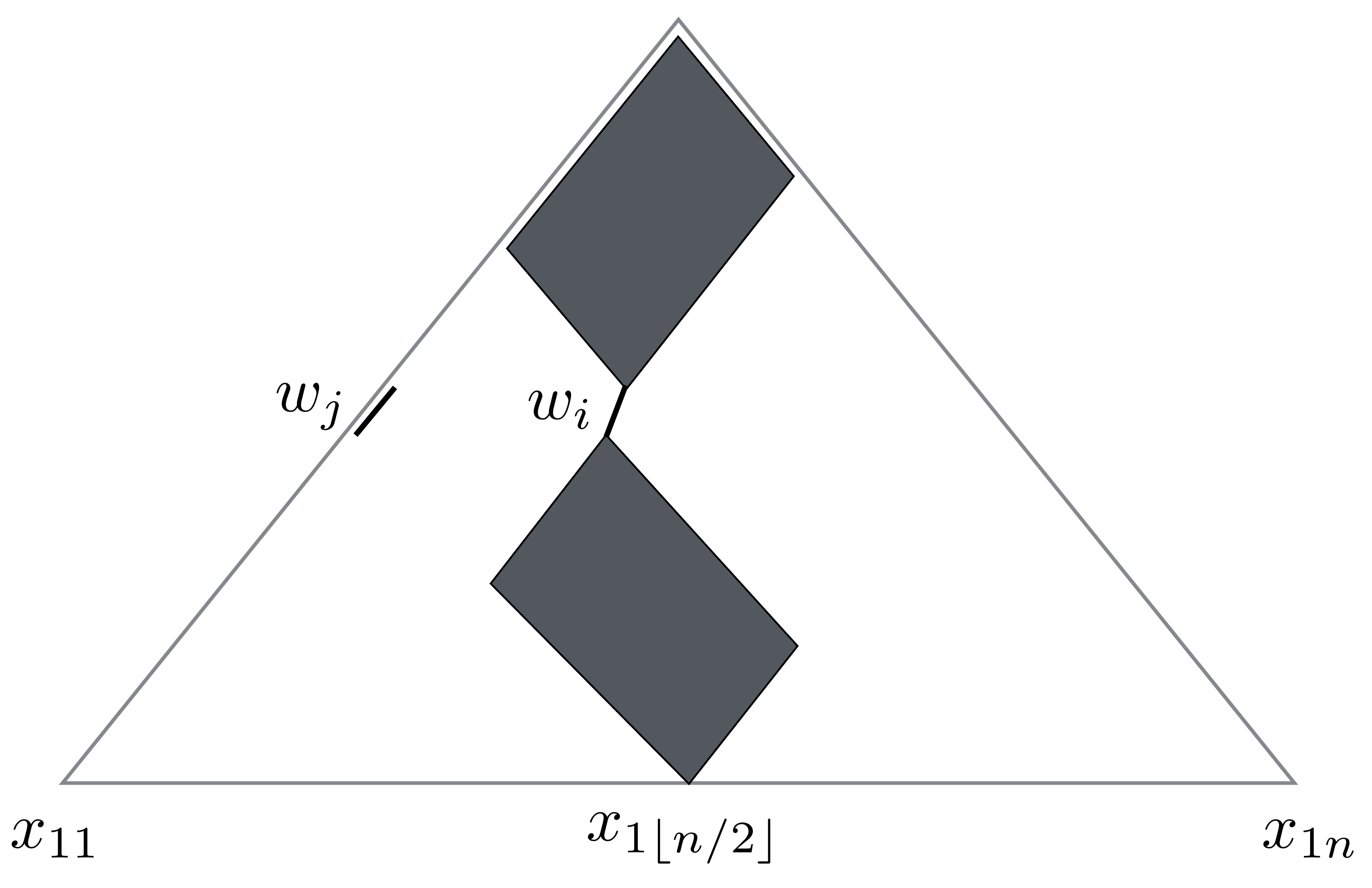}
	\caption{All paths through the grey area contributes to the gradient of $w_i$, while only the path along the edge contributes to $w_j$ (assuming there is no path from $x_{1\lfloor n/2 \rfloor}$ to $w_j$).}
	\label{fig_paths_for_a_weight}
\end{figure}

We can use this approach to compare the gradients of the cost function $c$ with respect to different $w_i$.
The gradient sum over the two possible inputs $x_{11}=\pm 1$ is
\begin{align}
\sum_{x_{11}=\pm 1}\frac{\partial c}{\partial w_i} &= \sum_{x_{11}=\pm 1}\frac{\partial c}{\partial f} \frac{df}{d x_{m1}} \frac{\partial x_{m1}}{\partial w_i}
\end{align}
It is easy to see that 
\begin{align}
	\frac{\partial c}{\partial f} \left. \frac{df}{d x_{m1}}\right\rvert_{x_{11}=1}=-\frac{\partial c}{\partial f} \left. \frac{df}{d x_{m1}}\right\rvert_{x_{11}=-1}
\end{align}
and
\begin{align}
\frac{\partial x_{m1}}{\partial w_i} = \sum_{p \in S_i} \prod_{w_j \in p, j\neq i} w_j x_{11}
\end{align}
where $S_i$ is the set of paths which go through $w_i$ and one of $x_{11}$ and $x_{1\lfloor n/2 \rfloor}$.
Substitute these in, we have
\begin{align}
\label{eq:grad_final}
\sum_{x_{11}=\pm 1}\frac{\partial c}{\partial w_i} =2 \frac{\partial c}{\partial f} \left. \frac{df}{d x_{m1}}\right\rvert_{x_{11}=1} \sum_{p \in S_i} \prod_{w_j \in p, j\neq i} w_j
\end{align}
Since the first two terms on the r.h.s are the same when we compute $\frac{\partial c}{\partial w_i}$ for different $w_i$ in the same network, to compare the gradients we only need to consider the last term.
By using Observation~\ref{obs_variance_prop_num_paths}, we know that just after the initialization (or when $w_i$ have not been too correlated), the gradient of the weight that involved in more paths connecting $x_{11}$ or $x_{1\lfloor n/2 \rfloor}$ to $x_{m1}$ has a larger variance, and obviously the mean of the gradients are $0$.
On a speculation level,  this means most weights connecting $x_{1\lfloor n/2 \rfloor}$ to $x_{m1}$ changes faster compared to those connects $x_{11}$ to $x_{m1}$ in the same layer. Intuitively, this trends will continue, since the gradient w.r.t to a weight will be larger if other weights on the path have larger absolute values.
Indeed, in \autoref{fig:gradient_during_training}, we can see for one particular setting, the gradient with respect to a weight in the center region is in general larger compared to one in the corner throughout the whole training process.

There is another important factor that causes $x_{1\lfloor n/2 \rfloor}$ to have a larger influence on the output.
\begin{observation}
	If we assume that at the end of the training, all the weights have the same value, e.g. 1.
	In this case, $x_{1\lfloor n/2 \rfloor}$ will still have a larger coefficient in the expansion of the output, because there are more paths connecting from it.
	\label{obv_more_paths_at_end}
\end{observation}
So the high chance of $x_{1\lfloor n/2 \rfloor}$ having larger influence is likely a combined result of the Observation~\ref{obv_more_paths_at_end} and the gradient mechanism we discussed previously.

Now let us discuss the simplification we made above.
If the network is convolutional, which means the weights are shared in the same layer, Observation~\ref{obv_more_paths_at_end} is still true.
However, it is not clear whether the above argument about gradients is still relevant, as apparently the weights are changing at the same rates regardless of its position in one layer.
Below we will look at a very specific scenario, in which the magnitudes of the gradients with respect to different weights still play a role in the final outcome.
First, note that Observation~\ref{obs_variance_prop_num_paths} still holds, because it remains true that $\operatorname{Cov}(W_{p_1},W_{p_2})=0$ for any two different paths $p_1$ and $p_2$.
Assume in the Fig~\ref{fig_paths_for_a_weight}, $w_i$ and $w_j$ are shared weights and thus $w_i=w_j$.
Let us consider the possible scenario that the gradient with respect to $w_i$ and $w_j$ have different signs, but the norm of the gradient w.r.t $w_i$ is much larger.
Then the update of $w_i$ (and $w_j$) will be according to the gradient of $w_i$, which increase the likelihood of the final decision $x_{m1}$ to be based on $x_{1\lfloor n/2 \rfloor}$.
Overall, more study needs to be done regarding the convolutional neural networks.

The effect of activation function in the middle of the network will depend on the choice of the particular function and the initial distribution of the weights.
For example, if we initialize all weights with a small enough variance, then initially all $x_{ij}$ are still in the linear regime of activation function such as sigmoid or $\tanh$.
For the popular \textsc{relu} activation function (i.e. $y=\max (x,0)$), since it is linear whenever the input is larger then $0$, we can expect observation~\ref{obs_variance_prop_num_paths} to still be true.

Lastly, we considered a very simple input probability distribution. In general, the inputs are likely to be noisy, and the input-output relation is more complicated than an identity function. To have an toy example of noisy inputs, later in the next subsection we are going to perform a numerical experiment where we replace the $0$ input bits to a random $\pm 1$ in the input distribution we considered above. Intuitively, we can still argue that for $x_{ij}$ in the middle of a layer will have a larger signal-to-noise ratio compared to the ones on the edge, as a larger part of the variance come actually from the signal. This might suggest that the weights in the middle will change faster.

\subsection{Numerical results}

\begin{figure}
	\centering
	\includegraphics[width=\linewidth]{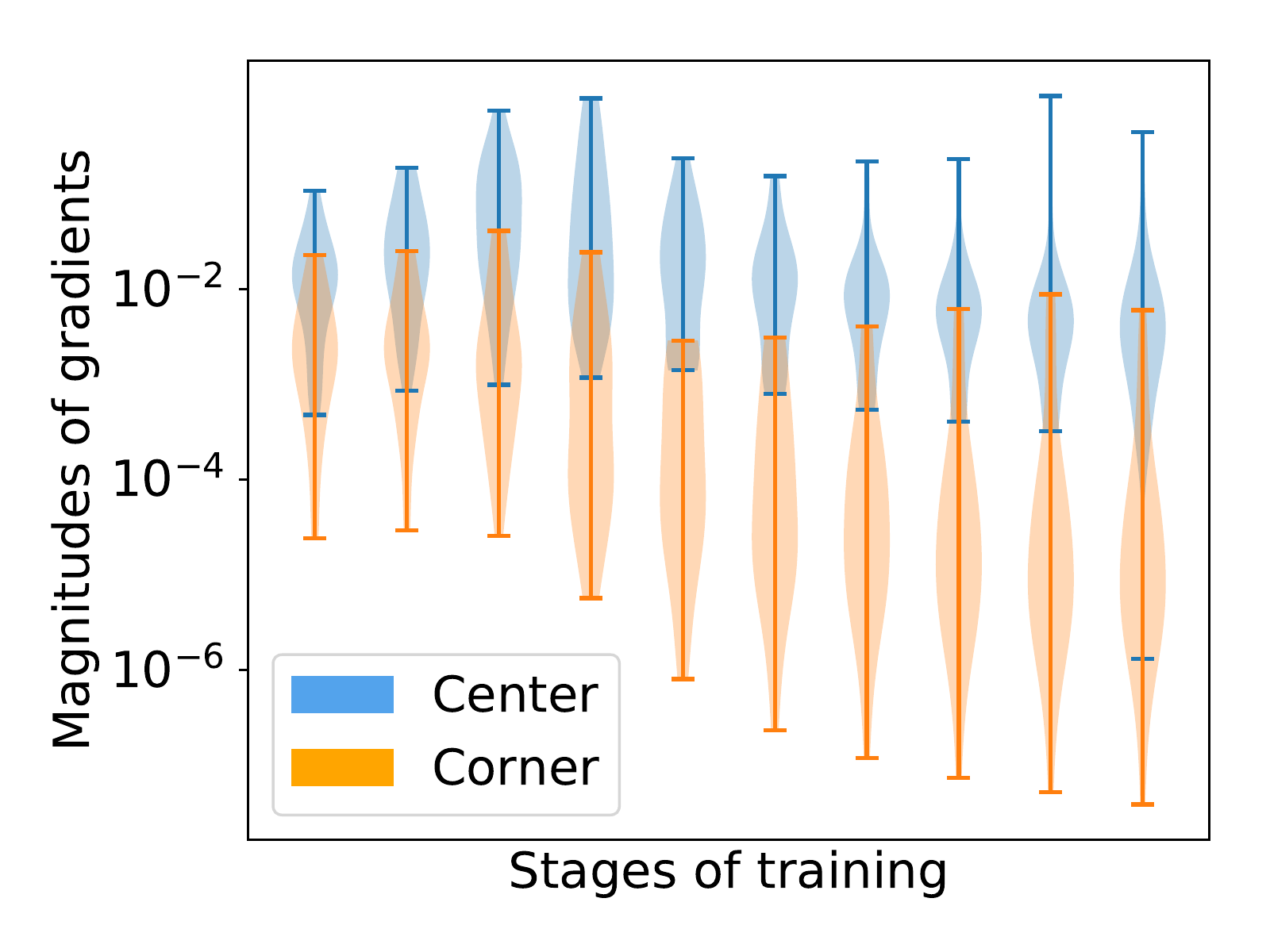}
	\caption{Absolute values of gradients with respect to two weights during different stages of the training. Both weights are connecting from the first to the second hidden layers. The data comes from running the same training process for 30 times, and we compute the gradient at 10 specific time steps during the training. In the plot, the vertical segments made of solid lines represent the ranges of the gradients, while the width of the transparent color patches represents the distribution. We can see the gradient of the weight in the center region is in general larger compared to the one in the corner.}
	\label{fig:gradient_during_training}
\end{figure}

We setup the network and the training data according to the description above, with the only change being that inputs are 2D with size $7\times 7$ and $9\times 9$.
In particular, the neural networks are not convolutional, but have the same local connectivity as a 2D convolutional one.
The number of channels for each hidden layer is $10$. We choose one of the corner input bits to be the same (or reverse) of the center bit. The weights of the network are initialized with normal distribution that has mean $0$ and standard variance $0.1$. We use the vanilla gradient descent with some exponential decay of the learning rate. The final output of the $\tanh$ is rounded to integer and compared to the desired output. The training stops when the average of cost function $c$ is smaller then $0.02$ for a batch of 100 training data (the threshold of $0.02$ is chosen so that when given the reversed input, the output of $\tanh$ will be rounded to $\pm 1$ instead of $0$). We then test the network on inputs where the chosen corner bit and the center bit have different signs.
We repeat the above procedure for 20 times on both $7\times 7$ and $9\times 9$ inputs. In all experiments, the network always output the center input bit when the center and corner input bits have different sign.
To verify the argument in the previous subsection that the gradients with respect to the variables in the center region are larger, we numerically compute the gradients at 10 selected time steps during the training process.
The result can be found in \autoref{fig:gradient_during_training}.
Here we set the input size to be $9\times 9$, and the number of channels for each hidden layer to be $1$.
We compute the gradient of the loss function with respect to two weights.
Both weights are connecting from the first to the second hidden layers.
One of the weights located in the center, while the other located in the corner corresponding to the non-zero element in the input.
In the plot, each vertical segment represents the range of the gradients at a particular time step of the 30 training runs, while the width of transparent color patches represents the distribution.
Indeed we can see the gradient of the weight in the center region is in general larger compared to the one in the corner.

\begin{figure}
	\centering
	\includegraphics[width=0.9\linewidth]{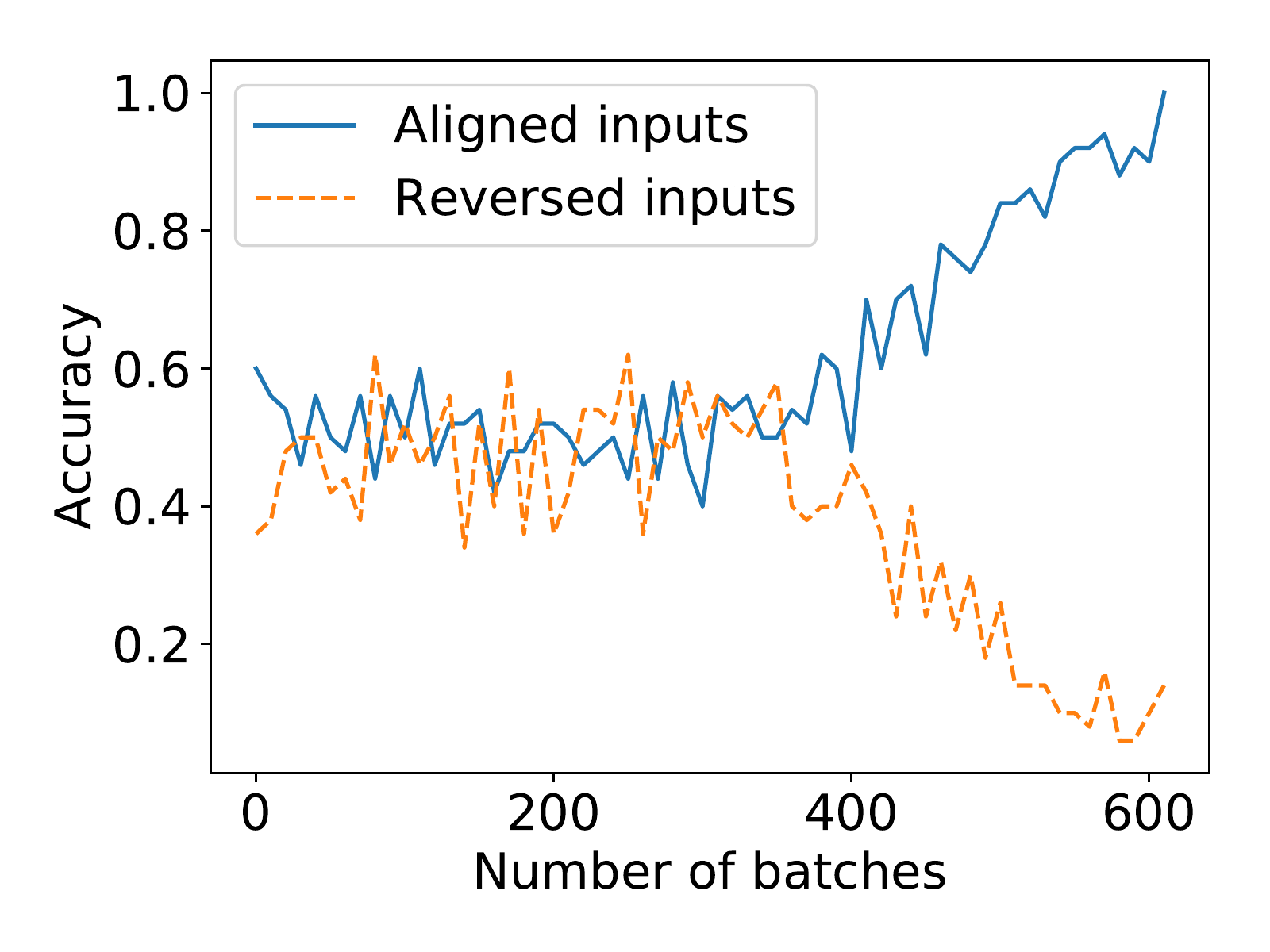}
	\caption{Accuracy of matching the center input bits during training. Inputs are noisy and there are non-linear activation functions for all layers.
	The blue solid line corresponds to the accuracy when given aligned inputs, while the orange dotted one corresponds to reversed inputs. Note that in order to differentiate the lines, the accuracy for reversed inputs is defined to be the percentage of output matching the corner bit. Thus towards the end of the training, the outputs almost always matching the center bits when given reversed input. The accuracy is evaluated on a batch of 50 inputs.}
	\label{fig:aligned_vs_reversed}
\end{figure}

We also do a test where the scenario is more complicated (and to some degree more practical). As discussed previously, we can add noise by changing all the $0$ in the inputs to a random $\pm 1$. \textsc{relu} activation function (i.e. $y=\max (x,0)$ is added to the middle layers of the network. The size of the input is $9\times 9$, and accuracy is evaluated by comparing the signs of the final $\tanh$ to the desired outputs. 
Otherwise the setting is the same as in the previous experiment.
In \autoref{fig:aligned_vs_reversed}, we plot the accuracy during the training process. We can see while the accuracy rises when given the aligned inputs (i.e. the particular corner bit and the center bit are equal), the network becomes more deterministically at outputting the center bit when given reversed inputs (i.e. the particular corner bit and the center bit have inversed sign).
The test are repeated several times where similar results are observed.

